%% file: main_preprint.tex
\documentclass[twocolumn]{article}  

\usepackage{graphicx}
\usepackage[T1]{fontenc}
\usepackage[subrefformat=parens]{subcaption}

\makeatletter
\let\cl@chapter\undefined
\makeatletter

\usepackage{amsmath,amssymb}   
\usepackage{amsthm} 
\usepackage{mathtools} 
\usepackage[numbers,sort&compress]{natbib}
\usepackage{hyperref}
\hypersetup{colorlinks=true,pdfborder={0 0 0}}
\usepackage{cleveref}  
\usepackage{amsbsy}
\usepackage{bm}
\usepackage{algorithmic}
\usepackage{algorithm}
\usepackage{url}
\usepackage{color}
\usepackage{multirow}
\usepackage{booktabs}
\usepackage{autobreak}
\allowdisplaybreaks 

\providecommand{\deltas}{\delta_{\mathrm{s}}}
\providecommand{\deltap}{\delta_{\mathrm{p}}}

\providecommand{\degree}{^{\circ}}
\providecommand{\subm }{_{\mathrm{m}}}
\providecommand{\vm}{v_{\mathrm{m}}}
\providecommand{\np}{n_{\mathrm{P}}}
\providecommand{\lpp}{L_{\mathrm{pp}}}

\providecommand{\hpr}{{_\mathrm{HPR}}}
\providecommand{\bt}{{_\mathrm{BT}}}
\providecommand{\train}{^{\text{(train)}}}
\providecommand{\valid}{^{\text{(validation)}}}
\providecommand{\test}{^{\text{(test)}}}
\providecommand{\ca}{^{\text{(CA)}}}
\providecommand{\simu}{^{\text{(sim)}}}
\providecommand{\opt}{_{\text{opt}}}
\providecommand{\Ltrainprime}{\mathcal{L}^{\prime \mathrm{(train)}}\opt}
\providecommand{\Lvalidprime}{\mathcal{L}^{\prime \mathrm{(validation)}}\opt}
\providecommand{\Ltestprime}{\mathcal{L}^{\prime \mathrm{(test)}}\opt}
\providecommand{\Lcaprime}{\mathcal{L}^{\prime \mathrm{(CA)}}\opt}

\providecommand{\argmin}{\operatornamewithlimits{argmin}} 
\providecommand{\R}{\mathbb{R}} 

\crefname{equation}{Eq.}{Eqs.}%
\crefname{figure}{Fig.}{Figs.}%
\Crefname{equation}{Eq.}{Eqs.}%
\Crefname{figure}{Fig.}{Figs.}%

\theoremstyle{definition}
\newtheorem{idea}{Idea}
\newtheorem{req}{Requirement}

\crefname{idea}{Idea}{Ideas}
\crefname{req}{Requirement}{Requirements}

\usepackage[normalem]{ulem}

\newcommand{\AddD}[1]{\textcolor{black}{#1}}	
\newcommand{\EraseD}[1]{\if0{#1}\fi}	

\begin{document}

\title{Development of a Mathematical Model for Harbor-Maneuvers to Realize Modeling Automation 
}

\author{Yoshiki Miyauchi$^{1}$         \and Youhei Akimoto$^{2,3}$ \and
        Naoya Umeda$^{1}$              \and Atsuo Maki$^{1}$
}

\date{%
\flushleft{\footnotesize
    $^1$Osaka University, 2-1 Yamadaoka, Suita, Osaka, Japan\\%
    $^2$Faculty of Engineering, Information and Systems, University of Tsukuba, 1-1-1 Tennodai, Tsukuba, Ibaraki 305-8573, Japan\\%
    $^3$RIKEN Center for Advanced Intelligence Project, 1-4-1 Nihonbashi, Chuo-ku, Tokyo 103-0027, Japan\\[2ex]%
    Email: yoshiki\_miyauchi@naoe.eng.osaka-u.ac.jp; maki@naoe.eng.osaka-u.ac.jp \\
}}



\maketitle

\begin{abstract}
A simulation environment of harbor maneuvers is critical for developing automatic berthing. Dynamic models are widely used to estimate harbor maneuvers. 
However, human decision-making and data analysis are necessary to derive, select, and identify the model because each actuator configuration needs an inherent mathematical expression. 
We proposed a new dynamic model for arbitrary configurations to overcome that issue. The new model is a hybrid model that combines the simplicity of the derivation of the Taylor expansion and the high degree of freedom of the MMG low-speed maneuvering model. 
We also developed a method to select mathematical expressions for the proposed model using system identification. Because the proposed model can easily derive mathematical expressions, we can generate multiple models simultaneously and choose the best one. This method can reduce the workload of model identification and selection.
Furthermore, the proposed method will enable the automatic generation of dynamic models because it can reduce human decision-making and data analysis for the model generation due to its less dependency on the knowledge of ship hydrodynamics and captive model test.
The proposed method was validated with free-running model tests and showed equivalent or better estimation performance than the conventional model generation method.
\end{abstract}

\section{Introduction}
 
   

Research and development of Maritime Autonomous Surface Ship (MASS) are active, including automatic berthing.
The development of controllers for MASS---including parameter tuning, verification, and certification by class society---, ultimately requires verification in a real ship. Still, a significant part of development in a simulation environment would be helpful from a cost and safety. A dynamic model of a ship's maneuver (i.e., maneuvering model or system-based mathematical model) is widely used to estimate ship maneuvering in numerical simulations from the standpoint of computation time.
 
The harbor\EraseD{-, approach- and berthing-} maneuvers of a ship \AddD{ is an inclusive maneuver that includes: leaving and entering the port, approach maneuvers to the berth, berthing, and also unberthing;} \AddD{hence it contains} various types of motions, i.e., forward, astern, acceleration and deceleration, turn, pivot-turn, stop, crabbing, position-keeping\EraseD{These motions are referred to as Complex Low-Speed maneuver or CLS maneuver}.  A dynamic model for harbor\EraseD{-, approach- and berthing-} maneuvers is generally more complex than that for \AddD{ocean} navigation because the flow field and usage of actuators are much more complex and because the ship is navigating close to obstacles, the high estimation accuracy is required.

Dynamic models for \EraseD{estimating CLS maneuvering}\AddD{harbor maneuvers} can be divided into two categories \AddD{how they address the complexity}; either a complex model that switches and combines sub-model with\EraseD{ the forward state and a large drift angle and propeller reversal} \AddD{various} condition, or a simpler one that uses unified mathematical expression and parameters for all state for practicality. The former method is represented by the low-speed maneuvering MMG model \citep[e.g., ][]{Miyauchi2022SI,Yasukawa2021}, and the latter is represented by Abkowitz model \EraseD{which takes}\AddD{with variable} propeller rotation \citep{Abkowitz1980} \EraseD{into account} \AddD{and Fossen's model} for dynamic positioning \citep{fossen2021handbook}.

\subsection{Problem Definition}
Generating a dynamic model of a particular ship generally consists of the following four tasks. Therefore, in this paper, the term ``model generation'' refers to the entire process that includes the following four tasks, and, a ``user'' is a person who generates a dynamic model and utilizes the dynamic model in a simulation environment.

\paragraph{Derivation of the model's mathematical expression.}
The user derives a new dynamic model if the existing dynamic model is not sufficient. The derivation here refers to the derivation of the mathematical expressions of the dynamic model. This step is unnecessary if the user thinks an existing dynamic model is sufficient.

\paragraph{Model selection.} 
The user selects the dynamic model to be used\EraseD{ based on the motion to be estimated and the configuration of the ship's actuators}. The selection is based on the findings of previous studies or results of maneuvering basin tests conducted by the user.

\paragraph{Parameter identification.}
The user obtains model parameters for the selected model from maneuvering basin tests, databases, or empirical formulae.

\paragraph{Validate dynamic model.}
The user performs a maneuvering simulation by reflecting the obtained model parameters in the dynamic model. The dynamic model is validated by comparing the simulation results with know maneuvers: a maneuvering trial of a ship or a free-running model ship.


\subsection{Difficulties of Existing Modeling Methods}\label{sec:RQ_vectwin_abko}


Let us assume that the user's goal is to build a maneuvering simulation environment that can simulate \EraseD{CLS maneuvering}\AddD{harbor maneuvers}. If the hull-form of the ship is fixed and the accuracy of maneuvering simulation is required, model parameters are often identified \AddD{by the captive model test \citep{Captive2017,Chislett1965,Yasukawa2015}. The standard method of the MMG model \citep{Yasukawa2015} also uses captive model tests to identify model parameters. Moreover, even for harbor maneuvers, several studied \citep{Shouji1992IFAC,Hasegawa1994,Maki2020_berthing1,Maki2020_berthing2,Miyauchi2021planning,RACHMAN2022111127,Shimizu2022,Akimoto2022_saddle} on berthing control have used captive model tests to identify model parameters. However,}
existing dynamic models and captive-model-tests-based scheme seems problematic because the user's workload to generate the model is large. In particular, the following are some examples:

\paragraph{Special testing facility.}
The model parameters are obtained from hydrodynamic forces measured in a towing tank or a maneuvering basin; hence, the model cannot be generated without using such a facility. Recently, Computational Fluid Dynamics (CFD) has been used \AddD{to alternate captive model tests \citep{Stern_SIMMAN,Guo2017,Sakamoto2019CFD}.} However, powerful computational resources and CFD expertise are required.
    
\paragraph{Cost of parameter identification}
The time and cost required for the experiment become non-negligible due to the increased number of test conditions caused by the complexity of the model of \EraseD{CLS maneuver}\AddD{harbor maneuvers}.
    
\paragraph{Selection of the dynamic model.}
     So far, several dynamic models have been proposed that are applicable to harbor maneuvers (see \Cref{sec:RW_vectwin_SI}), but none of them can be called the de-facto standard yet. Therefore, the user must select an appropriate model (and its mathematical expression) for their vessel of interest. Selection can be done by comparing various models with a known maneuver or comparing estimated hydrodynamic forces with captive model tests.

\paragraph{\AddD{Diversity of Actuator Configurations.}}
       \AddD{A ship's actuator system includes a wide variety of configurations, e.g., a single-propeller and single-rudder ship, a twin-propeller and twin-rudder ship, a ship with multiple azimuth thrusters, and so on. The mathematical expression of the dynamic model is unique to each actuator configuration; thus, the user needs to prepare the model and implement a numerical code for each actuator configuration.} 
\paragraph{Knowledge of Ship hydrodynamics and model test.}
        \AddD{The user must have appropriate knowledge of captive model tests, hydrodynamics on ship maneuvering, and the dynamic model itself to identify model parameters from captive model tests. However, for the users who aim to build automatic berthing systems, those knowledge is not necessarily their area of expertise. In such cases, a different person who is an expert in that area should be asked for support.}

\EraseD{In addition, the complexity of the MMG model makes the choice of model structure problematic. If the accuracy of motion estimation is important, a low-speed maneuvering version of the MMG model would be selected. MMG models are based on a modular structure, so each module, such as the hull, propeller, and rudder, has its own dynamic model. The model for each module estimates the open-water characteristics based on hydrodynamics. And finally, net force acting on the ship is modeled by adjust the model with interaction and modification coefficients to balance the results. Many models have been proposed for CLS maneuvering, therefore, the user must choose from a number of derived models, but there does not seem to be a de-facto standard model at this time. Thus users need to select a model structure by comparing it with hydrodynamic force measurements made by captive model tests and/or flow field measurements around the ship's hull that were used with particle image velocimetry.}





\subsection{Related Works}\label{sec:RW_vectwin_SI}
\AddD{Among the difficulties stated in \Cref{sec:RQ_vectwin_abko}, ``special testing facility'' and ``cost of parameter identification'' can address by the System Identification (SI) technique. On the SI for ship maneuvering, SI identifies model parameters from a set of ship trajectories. SI for ship maneuvering was first introduced by \"{A}str\"{o}m and K\"{o}llstr\"{o}m \citep{Astrom1976}, and Abkowitz \citep{Abkowitz1980} in the late 1970s and early 1980s, and then numerous studies have been done for both from full-scale\citep{Astrom1976,Abkowitz1980,Mei2019,Zhu2020} ship and model ship \citep{Araki2012a,Xie2020,Bai2019,Xu2020}.}

\AddD{However, research on SI have been conducted mainly on dynamic models for IMO standard maneuvers \citep{IMO_maneuver}, ocean voyage, and autopilot application. Only a few studies have been conducted on SI for dynamic models on harbor maneuvers. The authors previously studied \citep{Miyauchi2022SI} the feasibility of SI for low-speed maneuvering MMG model for single-propeller, single rudder ship. The MMG model used in the previous study \citep{Miyauchi2022SI} can represent characteristics of harbor maneuvers, e.g., large drift angle, thrust reduction on propeller reversal, and rudder characteristics dependency on propeller wake. Model parameters were identified from trajectories for the scaled free-running model. SI of low-speed maneuvering MMG model 
for single-propeller, single rudder ship was also done by Sawada et al. \citep{Sawada2020}. In \citep{Sawada2020}, they used full-scale ship trial data as train data and succeeded in developing a berthing controller with a simulation environment using the identified model. Using Fossen's model \cite{fossen2021handbook}, SI of an urban ferry with azimuth thrusters was done by \citep{Pedersen2019}.}


\AddD{All of the above previous research \citep{Miyauchi2022SI,Sawada2020,Pedersen2019} used dynamic models which categorized as the Hydrodynamic model \citep{Ogawa1978}. The Hydrodynamic model is a model that was derived and designed to identify the model parameters based on hydrodynamic assumption, observation, and measurement. The other kind of model is the Response model \citep{Ogawa1978}, which considers only the relationship between the ship’s control input and response, i.e., the motion of the ship. The most popular Response model would be Nomoto's KT model \citep{Nomoto1957En}. Meanwhile, application of neural networks (NN) as a Response model has been actively studied in the last two decades \citep{Moreira2003,Rajesh2008,Oskin2013}. Models using NN are highly capable of modeling the nonlinear dynamics; however, to the best of our knowledge, Wakita et al. \citep{Wakita2022} is the only case of SI using NN for harbor maneuvers.}



\subsection{Objective and Contribution of the study}
\AddD{We recognize that to overcome the remaining difficulties---selection of a dynamic model, diversity of actuator configurations, and knowledge of ship hydrodynamics---a new Response-model-type dynamic model is necessary. The hydrodynamic model has a favorable feature that the physical meaning of the formulae is understandable; therefore, flow field and forces can be observed from the model. However, it needs to derive model formulae with due consideration of hydrodynamics for each configuration. On the other hand, as a simulation environment for R\&D of automatic berthing, how the ship moves, i.e., the relationship between the control input and response, is important. Whether the dynamic model can represent the flow field and hydrodynamic forces is desirable but not mandatory. Hence, the response model, especially a model using NN, is an option to overcome those issues; however, behavior on the extrapolation region of train data is questionable because the model is not understandable and needs treatment to impose a constraint to be physically reasonable (e.g., \citep{Wakita2022}).}


\EraseD{Therefore, a dynamic model that does not require special testing facilities and model selection based on sufficient observation of the flow field to generate a dynamic model for each individual vessel, and that can be generated automatically once the user has input data, could facilitate research and development of automatic berthing and unberthing.
From the standpoint of facilitating R\&D, changing the structure of the dynamic model for each configuration of the ship's actuators would not necessarily be desirable. Each time the dynamic model is changed, or each time the configuration of the actuators is changed, a new program for the model to be identified would have to be coded.}

\AddD{Therefore, we aim to propose a new Response model whose formulae are easily derivable for arbitrary actuator configurations, have enough degree of freedom to express harbor maneuvers, and not having the downside of NN models. In addition, we aim to propose a method to select formulae of the proposed model which does not depend on captive model tests and knowledge of ship hydrodynamics. Our previous work \citep{Miyauchi2021Abkowitz} proposed a 4-quadrant Abkowitz model for single-propeller, single-rudder ships. This previous model was based on Taylor expansion, so the polynomials are easily derived. In addition, the previous model has four sets of model parameters to model the change of characteristics caused by propeller reversal and astern conditions. In this study, we expand the previous model \citep{Miyauchi2021Abkowitz} for arbitrary actuators and develop a method for model generation using the SI. By utilizing SI, we will be able to be less dependent on captive model tests and ship hydrodynamics in model generation. Furthermore, the proposed method enables the automatic generation of dynamic models because human intervention in  model derivation, selection, and identification can be reduced. Our previous literature \citep{Miyauchi2022SI_conference} described the initial results of the investigations done in this paper. This paper presents the results more extensively, with more details, and with several revisions.}

\EraseD{Based on the above problems, the objective of this study is to develop a new dynamic model for ship berthing and unberthing simulator. This study extends the 4-quadrant Abkowitz model \citep{Miyauchi2021Abkowitz} proposed earlier by the authors.}

The major contributions of this study are as follows:

\begin{enumerate}
\item We proposed a new dynamic model, from which the model formula is easily derived according to simple rules and is complex enough to handle complex \EraseD{ship handling at low speeds}\AddD{phenomena of harbor maneuvers}. The new model is a hybrid model that combines the simplicity of the derivation of the Abkowitz model, based on Taylor expansion, and the complexity of the MMG low-speed maneuvering model.
\item We proposed a method for selecting a model that does not require captive model tests \EraseD{ and sufficient observation of the flow field}\AddD{ and knowledge of ship hydrodynamics and modeling}\EraseD{ using the proposed model}. Because the proposed model can easily derive multiple formulae, we can generate multiple models using system identification and selects the model formulae that best fit the ship's trajectory. Therefore, we can reduce the effort of parameter identification and model selection for arbitrary actuator configurations.
\end{enumerate}

\section{Abkowitz-MMG hybrid model}\label{sec:model}
\subsection{Overview of the proposed model}

This section describes the dynamic model proposed in this study (hereafter referred to as the "proposed model"). The proposed model is a hybrid model that combines the simplicity of the Abkowitz model with the complexity of the MMG low-speed model. 
\paragraph{Requirements of the model.}
First, we describe the requirements of the proposed model and ideas of our approach to those requirements. The requirements of the proposed model were derived based on the issues raised in the introduction. The requirements are as follows.

\begin{req}\label{req:model_complexity}
    The model complexity can be increased easily. This allows for handling various levels of complexity of the flow field of harbor maneuvers.
\end{req}
\begin{req}\label{req:actuator}
    The user can easily derive the model's expression even if the configuration of the actuator of the subject ship changes.
\end{req}
\begin{req}\label{req:accuracy}
    The model must be accurate enough to evaluate the controller by maneuvering simulation, i.e., the same level of accuracy as existing methods.
\end{req}
\begin{req}\label{req:prop_reversal}
    The model can represent significant characteristic changes of fluid forces at berthing and unberthing (e.g., flow separation, propeller reversal).
\end{req}
\begin{req}\label{req:selection}
    Model selection and parameter identification of the proposed model does not require in-depth knowledge of dynamic models or detailed flow observation.
\end{req}
\begin{req}\label{req:cmt}
    The model does not require captive model tests through its generation process.
\end{req}
\begin{req}\label{req:stability}
    The model has robustness to input state and control input over the operation range of \EraseD{CLS maneuver}\AddD{harbor maneuvers}. In other words, the divergence of acceleration should not occur even for unknown data within the range of typical harbor maneuvers.
\end{req}

\paragraph{Basic ideas on the model design.}
We designed the proposed model to meet the above requirements with the following ideas.



\begin{idea}\label{idea:abko}
Following Abkowitz's polynomial model \citep{Abkowitz1964} (hereafter, Abkowitz model), the model is derived by Taylor expansion of the hydrodynamic forces. The complexity of the model can be easily increased by increasing the highest degree of the polynomial obtained by Taylor expansion or the number of terms employed (\Cref{req:model_complexity}). If the actuator configuration changes (\Cref{req:actuator}), we only need to add the actuator features (e.g., propeller revolution, rudder angle, azimuth thruster angle) to the Taylor expansion variables.
\end{idea}

\begin{idea}\label{idea:submodel}
  Following the concept of low-speed maneuvering MMG model, the model parameters are set to different values for each representative condition (e.g., propeller forward/reverse rotation). In addition, the same mathematical expression is maintained for all operating conditions, thereby preserving the simplicity (\Cref{req:selection,req:model_complexity,req:actuator}) while ensuring that the accuracy (\Cref{req:accuracy}) and the complexity (\Cref{req:prop_reversal}) could be achieved.
\end{idea}

\begin{idea}
Utilize the System Identification (SI) and use ship's trajectory (\AddD{i.e., the time }history of motion and actuator usage) as a training dataset to identify the model parameters. Suppose the user can easily derive several different mathematical expressions of the proposed model (\Cref{req:model_complexity}), the user can select a dynamic model by choosing the one that best fits the motion data for validation. This procedure allows the user to select the appropriate model without requiring observation of the flow field or knowledge of the model (\Cref{req:selection}). It also achieves captive model test free. 
\end{idea}

\begin{idea}\label{idea:stability}
The stability of the proposed model is not ensured because the model's expressions and parameter exploration domain are not based on hydrodynamics. Therefore, we intended to maintain the stability of the proposed model by the objective function and the dataset to be used in the optimization process. Details of the objective function and dataset are described in \Cref{sec:obj_func} and \Cref{sec:dataset}, respectively.
\end{idea}

\subsection{Derivation of Proposed Model}
In this section, we derive the mathematical expression of the proposed model based on the basic idea described in the previous section. First, the subject ship is a scaled model ship of a coastal vessel equipped with a single-propeller, VecTwin type twin-rudder, and bow thruster. The ship fixed coordinate system $\mathrm{O}-xy$ and space fixed coordinate system $\mathrm{O}_{0}-x_{0}y_{0}$ are defined as \cref{fig:coordinate systems} and equations of motion for the three degrees of freedom used in the standard MMG model \citep{Yasukawa2015} are:
\begin{figure}
    \centering
    \includegraphics[width=\linewidth]{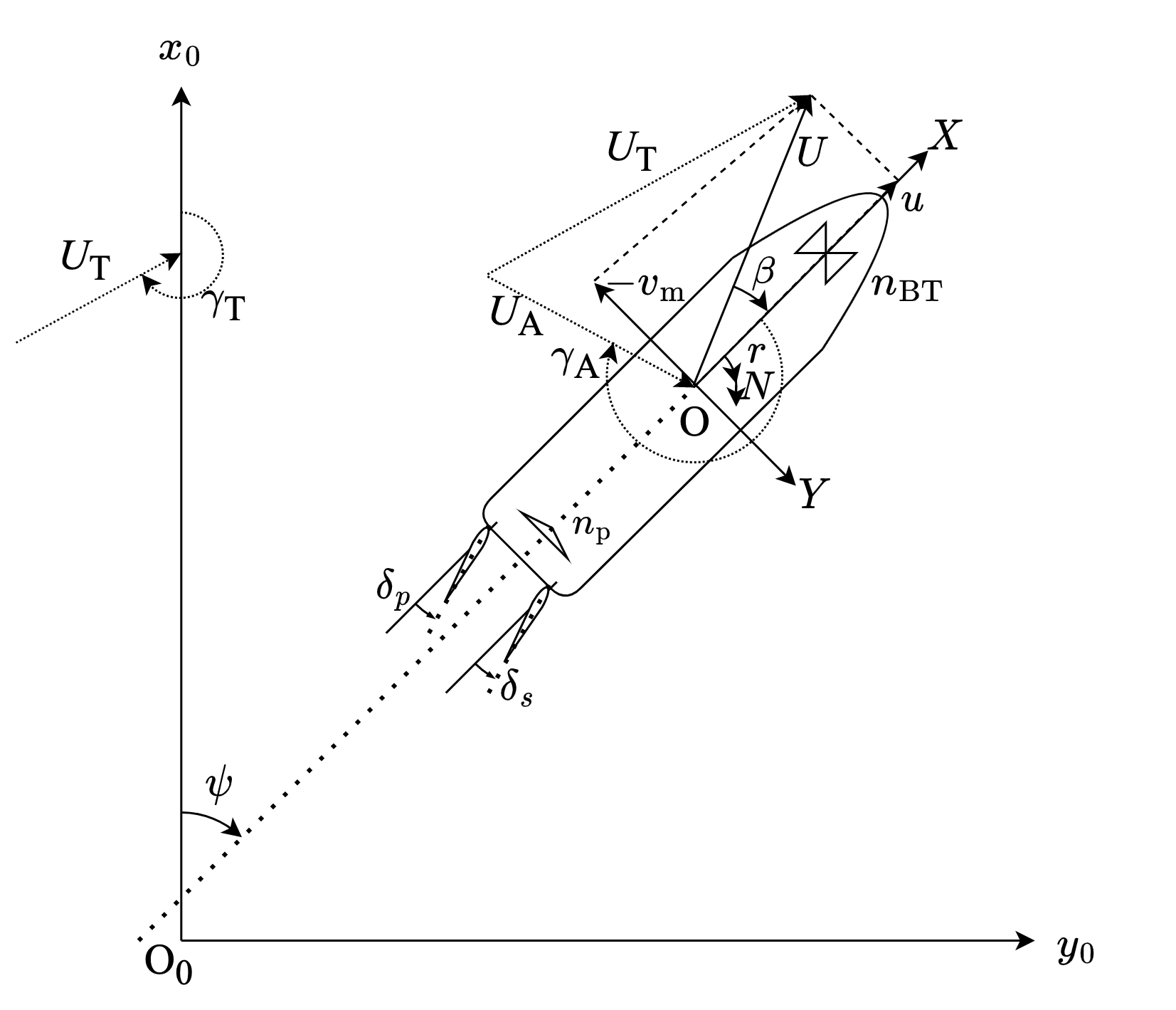}
    \caption{Coordinate systems.}
    \label{fig:coordinate systems}
\end{figure}
\begin{equation}\label{eq:motion}
    \begin{split}
        (m + m_x) \dot{u} - (m + m_y) v_m r -x_{G}mr^{2} &= X\\
        (m + m_y) \dot{v}_m + (m + m_x) u r +x_{G}m\dot{r} &= Y\\
        (I_{zz} + J_{zz}+x_{G}^{2}m) \dot{r} + x_G m (\dot{v}_m + ur) &= N \enspace .
        \end{split}
\end{equation}
\AddD{Here, notations for physical parameters are: $m$ is ship's mass, $I_{zz}$ is the moment of inertia at the center of gravity (CG), $m_x$ and $m_y$ are added mass, $J_{zz}$ is added inertia, $x_G$ is longitudinal distance to CG from midship, $X,~Y,~N$ are forces and moment acting on the body except added mass. Notations for kinematic variables are: $\psi$ is ship heading, $u$ and $\vm$ are longitudunal and lateral speed of motion on $O-xy$ system, $r$ is yaw angular velocity, $\beta$ is drift angle, and $\dot{(\cdot)}$ represent the time derivative. Notations for actuator features are: $\np$ and $n\bt$ are revolution numbers of propeller and bow thruster, $\deltas,~\deltap$ are rudder angles of starboard and port side rudder. Notations for wind are: $U_{\mathrm{T}}$ and $\gamma_{\mathrm{T}}$ are true wind speed and direction, $U_{\mathrm{A}}$ and $\gamma_{\mathrm{A}}$ are apparent wind speed and direction, respectively.}

Next, the right-hand side of \Cref{eq:motion} is expressed as a polynomial by Taylor expansion. The classical Abkowitz model assumes the perturbation motion of a ship from a certain forward speed $u=U_{0},~v=0,~r=0$. Then, expand the hydrodynamic forces or their dimensionless forms by the state variables $\boldsymbol{s}_{t}$, the control vectors $~\boldsymbol{a}_{t}~$ and their time derivatives $\dot{\boldsymbol{s}}_{t},~\dot{\boldsymbol{a}}_{t}$. 
To apply Abkowitz model to \EraseD{CLS maneuver}\AddD{harbor maneuvers}, we use Taylor expansion with ideas below:
\begin{idea}
    \AddD{Assume that hydrodynamic forces and moments are function of $\boldsymbol{s}_{t}$, $\boldsymbol{a}_{t}$ and added mass component on left hand side of \Cref{eq:motion}. The users can include other added mass components and $\dot{\boldsymbol{a}}_{t}$ in the dynamic model if necessary, however generally those are neglected in MMG model.}
\end{idea}
\begin{idea}
    No scale conversion is performed since we assumed that the user estimates the dynamic model directly from the motion of a full-scale ship. This means that the hydrodynamic forces are expanded by Taylor expansion without non-dimensionalization. This is also because to avoid zero division caused by the non-dimensionalization\footnote{$X^{\prime} = X /(0.5 \rho LdU^{2}),~Y^{\prime} = Y /(0.5 \rho LdU^{2}),~N^{\prime} = N /(0.5 \rho L^{2}) dU^{2})$} of the hydrodynamic forces at $U=0$.
\end{idea}
\begin{idea}
    Assuming slow-speed motion, hydrodynamic forces are expanded around the origin $(\boldsymbol{s}_{t}=0,~\boldsymbol{a}_{t}=0)$ by Taylor's expansion.
\end{idea}
\begin{idea}\label{idea:decompose_wind}
    Hydrodynamic forces are decomposed into the above-water (wind) and below-water (underwater) contributions to account for the wind force, which is relatively important in harbor maneuvers. This decomposition is appropriate because the relative wind velocity governs the wind force while ship's state and control input governs the underwater contribution. While the underwater contribution is Taylor expanded,the wind force is modeled using an existing wind force model.
\end{idea}
Accordingly, \AddD{let us start }the derivation of the dynamic model from the decomposition of the hydrodynamic forces on the right-hand side of the \Cref{idea:decompose_wind} into wind pressure forces $X_{\mathrm{A}},~ Y_{\mathrm{A}},~N_{\mathrm{A}}$ and other hydrodynamic forces. Next, \AddD{let us focus on}\EraseD{we consider} hydrodynamic forces acting below the water surface. Since the subject ship has a bow thruster, the following is assumed.
\begin{itemize}
    \item Bow thruster force $~Y\bt,~N\bt$ is separated with the hydrodynamic forces of the hull+rudder+propeller $X\hpr,~Y\hpr,~N\hpr$, because bow thruster is located at the bow, away from the rudder and propeller, so we assume that the influence of the flow interaction between bow thruster and  rudder/propeller is small. If this were a stern thruster, the coupling with the propeller and rudder would need to be considered.
    \item Bow thruster is assumed that it does not generate force in the X direction.
\end{itemize}
From these assumptions, we yield the following:
\begin{align}\label{eq:decompose}
    X &= X\hpr + X_{\mathrm{A}} \\
    Y &= Y\hpr + Y_{\mathrm{BT}}+ Y_{\mathrm{A}} \\
    N &= N\hpr + N_{\mathrm{BT}}+ N_{\mathrm{A}} \enspace.
\end{align}
The wind force $X_{\mathrm{A}},~ Y_{\mathrm{A}},~N_{\mathrm{A}}$ is modeled by exsiting polynomial-type model (see \Cref{sec:vecwtin_si_wind}) proposed by \citep{Fujiwara1998}. The coefficients of the wind force model were optimized simultaneously with the coefficients of $X\hpr,~Y\hpr,~N\hpr$ by SI. As a first step of research, bow thruster was neglected, although the subject ship is equipped with a bow thruster; therefore, always $Y\bt=N\bt=0$.

Next, we select the variable vector $\boldsymbol{x}=(\boldsymbol{s}_{t},~\boldsymbol{a}_{t})$ for Taylor expansion of $X\hpr,~Y\hpr,~N\hpr$. We selected following variables for state variables vector $\boldsymbol{s}_{t}$ and control input vector $\boldsymbol{a}_{t}$, where the subscript $t$ is the value at time $t$:
\begin{align}
    \boldsymbol{s}_{t} &\equiv(u,~v\subm ,~r) \in \R^{3} \\
    \boldsymbol{a}_{t} &\equiv(\np,~\sin(\deltas),~\cos(\deltas),~\sin(\deltap),~\cos(\deltap)) \in \R^{5} \enspace.
\end{align}
Here, we use trigonometric functions of the rudder angles $\deltas,~\deltap$ for the port and starboard rudder features. Most Abkowitz models use the rudder angle $\delta$, but we must consider that large rudder angles are used in harbor maneuvers.


If the user wants to derive the dynamic model for other ship configurations, add appropriate features to $\boldsymbol{a}_{t}$. For example, for a ship equipped with a stern thruster, the revolution number of the stern thruster can be added to $\boldsymbol{a}_{t}$. Likewise, the revolution number of the azimuth thruster and the $\sin$ and $\cos$ functions of the operating angle can be used for an azimuth thruster.

\subsubsection{Derivation of Mathematical expressions by Taylor expansion}
Taylor expansion is now performed for $X\hpr,~Y\hpr$, and $N\hpr$. Hereafter we only show the manipulation for $X\hpr$, but $Y\hpr$ and $N\hpr$ are also derived similarly. $X\hpr$ is expanded for $\boldsymbol{x}$ as follows:
\begin{align}\label{eq:Taylor}
    \begin{aligned}
    X&\hpr(x_{1},~\ldots,~ x_{k}) \\
    &=   \left.X\hpr(x_{1},~\ldots,~ x_{k})\right|_{\boldsymbol{x}=0}  \\
    &+\frac{1}{1!}\left(\sum_{i=1}^{k}x_{i}\left.\frac{\partial}{\partial x_{i}}\right|_{\boldsymbol{x}=0}\right)X\hpr(x_{1},~\ldots,~ x_{k}) \\
    &+\left.\frac{1}{2!}\left(\sum_{i=1}^{k}x_{i}\frac{\partial}{\partial x_{i}}\right|_{\boldsymbol{x}=0}\right)^{2}X\hpr(x_{1},~\ldots,~ x_{k}) \\
    &+\left.\frac{1}{3!}\left(\sum_{i=1}^{k}x_{i}\frac{\partial}{\partial x_{i}}\right|_{\boldsymbol{x}=0}\right)^{3}X\hpr(x_{1},~\ldots,~ x_{k}) \\
    &+ \cdots \enspace,
    \end{aligned}
\end{align}
where
\begin{equation}
\begin{split}
    \boldsymbol{x} = (&x_{1},~\ldots,~ x_{k}) \\
    = \big(&u,~v\subm ,~r,~\np,~\sin(\deltas),~\cos(\deltas),~\\
    &\sin(\deltap),~\cos(\deltap)\big) \in \R^{8} \enspace .
\end{split}
\end{equation}
Ignoring the fluid memory effect, we can assume that $\left.X\hpr(x_{1},~\ldots,~ x_{k})\right|_{\boldsymbol{x}=0}=0$. Then the first term on the right-hand side of the \Cref{eq:Taylor} can be ignored. Next, expanding \Cref{eq:Taylor}, we obtain,
\begin{align}\label{eq:Taylor2}
    \begin{aligned}
    &X\hpr(\boldsymbol{s}_{t},\boldsymbol{a}_{t})\\
    &= \left.\frac{\partial X\hpr}{\partial u}\right|_{\boldsymbol{x}=0}u + \left.\frac{\partial X\hpr}{\partial \vm} \right|_{\boldsymbol{x}=0}\vm + \cdots \\
    &+\frac{1}{2!}\left(\left.\frac{\partial^{2} X\hpr}{\partial u^{2}}\right|_{\boldsymbol{x}=0}u^2 + \ldots
    \left.\frac{\partial^{2} X\hpr}{\partial \cos^{2}\deltap} \right|_{\boldsymbol{x}=0}\cos^{2}\deltap \right)\\
    &+\frac{1}{3!}\left(\left.\frac{\partial^{3} X\hpr}{\partial u^{3}}\right|_{\boldsymbol{x}=0}u^3 +\ldots
    + \left.\frac{\partial^{3} X\hpr}{\partial \cos^{3}\deltap} \right|_{\boldsymbol{x}=0}\cos^{3}\deltap \right)\\
    &+ \cdots \enspace.
    \end{aligned}
\end{align}
Partial derivatives of \cref{eq:Taylor2} are called hydrodynamic coefficients on Abkowitz model. By using common expressions of hydrodynamic coefficients, we can formulate \Cref{eq:Taylor2} as follows, 
\begin{equation}\label{eq:x_all}
    \begin{split}
        X&\hpr(\boldsymbol{s}_{t},\boldsymbol{a}_{t}) \\
        &= X_{u}u + \ldots + X_{\cos(\deltap)}\cos(\deltap) \\
        &+X_{uu}u^2 + \ldots + X_{\cos(\deltap)\cos(\deltap)}\cos^2(\deltap) \\
        &+X_{uuu}u^3  + \ldots + X_{\cos(\deltap)\cos(\deltap)\cos(\deltap)}\cos^3(\deltap) \\
        &+ \cdots \enspace,
    \end{split}
\end{equation}
where
\begin{align}
\begin{split}
    X_{x_{i}} & \equiv \left.\frac{\partial X\hpr}{\partial x_{i}}\right|_{\boldsymbol{x}=0} \\
    X_{x_{i}x_{j}} & \equiv \frac{1}{2!}\left.\frac{\partial^{2} X\hpr}{\partial x_{i}\partial x_{j}}\right|_{\boldsymbol{x}=0} \\
    X_{x_{i}x_{j}x_{k}} & \equiv \frac{1}{3!}\left.\frac{\partial^{3} X\hpr}{\partial x_{i}\partial x_{j}\partial x_{k}}\right|_{\boldsymbol{x}=0} \enspace.
\end{split}
\end{align}

\subsubsection{Selection of the term of polynomial equation}\label{sec:term_selection}
For polynomial-type dynamic models, the question is which terms in \Cref{eq:x_all} should be included in the polynomial. For existing dynamic models, the choice of formulae has been determined by comparison with the results of captive model tests. For example, for of MMG, see\citep{Matsumoto1980,Kose1982}, and for Abkowitz model see \citep{Chislett1965}. However, in this study, we assume a captive model test-free dynamic model (\Cref{req:cmt}). Therefore, the proposed model's order and term selection methods are as follows.
\begin{idea}\label{idea:select_term}
All terms are used unless they are obviously unnecessary or need modification.
\end{idea}
\begin{idea}
The highest order of the polynomials is defined by the capability of the optimization method and the user's computational resources. The user derives and compares several model formulae with different maximum order. In this study, second- and third-order models were used.
\end{idea}

The reasons are as follows. The available duration of the test facility for the captive model test-based method bounds the number of identifiable parameters. For example, the original Abkowitz model (\citep{Abkowitz1964}) used 65 parameters, and studies by the authors using the MMG low-speed maneuvering model (\citep{Miyauchi2022SI}) used 52 parameters to model the forces acting below the surface of the water. The authors believe that about 100 is a practical upper limit for the number of model parameters. To ensure accuracy with fewer terms, the terms employed should be examined based on theoretical considerations and observations on the flow field.
On the other hand, SI can increase the number of parameters as far as the performance of the optimization method allows. For example, CMA-ES used in this study (described in detail in\Cref{sec:cma}) can optimize up to several hundred dimensions. The proposed model is assumed to be a model that does not require detailed observation of the flow field; thus, terms are removed or modified only for obvious ones.

Based on the principles discussed above, we perform modification of \Cref{eq:x_all}. The proposed model is modified with the following three manipulations.

\paragraph{Modification A: Selection based on the symmetricity of the ship.}
    The term is selected based on the fact that the ship is symmetrical at the longitudinal center line of the hull. Original Abkowitz model \citep{Abkowitz1964} assumed that for a single-propeller, single-rudder ship; hence, $X$ is assumed as an even function of $\vm$\footnote{Precisely, \citep{Abkowitz1964} use the speed at the center of gravity, $v_{\mathrm{G }}$}, $~r,~\delta$, hence the odd powers of $\vm,~r,~\delta$ were deleted from the polynomial for $X$. Similarly, because $Y$ and $N$ were assumed as an odd function of $\vm,~r,~\delta$ due to symmetry, their even powers were deleted. 
    
    The proposed model also uses this symmetry assumption. We assume that $X$ is an even function of $\boldsymbol{x}_{\text{asym}}=\{\vm,~r,~\sin(\deltas),~\sin(\deltap)\}$, and remove the odd power terms $\sum_{i=1,3,\cdots}\{\vm+~r+~\sin(\deltas)+~\sin(\deltap)\}^{i}$. Also, from the polynomial of $Y$ and $N$, the $0$th power and even power terms $\sum_{i =0,2,\cdots}\{\vm+~r+~\sin(\deltas)+~\sin(\deltap)\}^{i}$ were removed. The asymmetric motion vector $\boldsymbol{x}_{\text{asym}}$ was selected for motions outside the symmetry plane of the hull.
    
    However, strictly speaking, $Y$ and $N$ are asymmetric and hence not purely odd functions of $\boldsymbol{x}_{\text{asym}}$. For example, for a single-propeller ship, turning forces of the port and starboard sides are different due to the rotational flow of propeller. In addition, significant lateral force and yaw moment will occur during propeller reversal of a single-fixed-pitch propeller ship \citep{Wagner1972,Fujino1978,Yoshimura1978}. That asymmetricity can be modeled by not removing the 0th powered term of $\boldsymbol{x}_{\text{asym}}$ (e.g., $Y_{u}, Y_{uu},\cdots$) \citep{Abkowitz1964}. However, we decided not to add terms to express asymmetricity to prioritize the simplicity of the model derivation.
    
\paragraph{Modification B: Delete terms containing $\sin^{2}$ and $\cos^{2}$.}
    Because $\sin^{2}+\cos^{2}=1$ duplicates other terms. For example,
    \begin{equation}\label{eq:example_modB}
    \begin{split}X_{u}u&+X_{u\sin^{2}\deltas}u\sin^{2}\deltas+X_{u\cos^{2}\deltas}u\cos^{2}\deltas \\
        &= (X_{u}+X_{u\sin^{2}\deltas}+X_{u\cos^{2}\deltas})u
    \end{split} \enspace.
    \end{equation}
    All the terms in parentheses in \Cref{eq:example_modB} are constants; hence those are unnecessary and must be deleted.
    
\paragraph{Modification C: Replace squared variables with modulus function.}
    Replace the squared variable, both on the 2nd order term and 3rd order term, with modulus function. For the $X$ polynomial, replace $x_{k}^{2}$ for $x_{k}\notin\boldsymbol{x}_{\text{asym}}$ to $x_{k}|x_{k}|$ and similarly in the $Y,~N$ polynomial, $x_{k}\in\boldsymbol{x}_{\text{asym}}$, let $x_{k}^{2}$ be $x_{k}|x_{k}|$. However, this is not done for odd power terms of non-coupling terms such as $uuu$ and $\vm\vm\vm$.

The results of the above operations are shown below. Here let us express the obtained polynomial for 3rd order model as the inner product of hydrodynamic coefficient vector $\boldsymbol{X}^{(i)}\in\R^{m} ,~\boldsymbol{Y}^{(i)}\in\R^{n}$, or $\boldsymbol{N}^{(i)}\in\R^{n}$ and even powers vector $\boldsymbol{z}_{\text{even}}\in\R^{m}$ of $\boldsymbol{x}_{\text{asym}}$ or odd powers vector $\boldsymbol{z}_{\text{odd}} \in\R^{n}$, as follows: 
\begin{align}\label{eq:innner_product}
    \begin{aligned}
        X\hpr &=  \boldsymbol{X}^{(i)} \cdot  \boldsymbol{z}_{\text{even}}\\
        Y\hpr &=  \boldsymbol{Y}^{(i)}   \cdot \boldsymbol{z}_{\text{odd}}\\
        N\hpr &=  \boldsymbol{N}^{(i)}  \cdot \boldsymbol{z}_{\text{odd}} \enspace, \\ 
    \end{aligned}    
\end{align}
where

\begin{align}\label{eq:xderivative}
    \begin{autobreak}
    \MoveEqLeft
        \input{xderivative.tex}
    \end{autobreak}
\end{align}

\begin{align}\label{eq:positive_vector}
    \begin{autobreak}
        \MoveEqLeft
        \input{z_even.tex}
    \end{autobreak}
\end{align}

\begin{align}\label{eq:yderivative}
    \begin{autobreak}
        \MoveEqLeft
        \input{yderivative.tex}
    \end{autobreak}
\end{align}

\begin{align}\label{eq:negative_vector}
    \begin{autobreak}
        \MoveEqLeft
        \input{z_odd.tex}
    \end{autobreak}
\end{align}
$N^{(i)}_{}$ is omitted because it is similar to \Cref{eq:yderivative}. 2nd-order model is consisted only with linear and 2nd order term of \Cref{eq:xderivative,eq:yderivative,eq:positive_vector,eq:negative_vector}. Here, the superscript $i$ represents the $i$-th parameter set of the coefficient vector Submodel, described in the following section.

\subsubsection{Submodel}\label{sec:submodel}
The proposed model uses the same polynomial formulae for all operational conditions but changes the values of the hydrodynamic coefficients for each condition, to achieve both simplicity of the derivation and complexity of the model, as described in \Cref{idea:submodel}.  The complexity of the model can be increased arbitrarily by increasing the number of conditional branches.
Here, we named the value set of coefficients for each condition as ``Submodel''.
 In our previous study \citep{Miyauchi2021Abkowitz}, a dynamic model of a single-propeller, single-rudder ship was developed with a Submodel for each propeller operating quadrant (defined by $u, n_{p}$).
Since the subject ship of the current study has a VecTwin rudder, the propeller is always in forward rotation. Therefore, the simplest Submodel would be Submodel-2, which switches by the sign of $u$. The conditional branch of Submodel-2 is shown in \Cref{alg:submod2}.
\begin{algorithm}[b]
    \caption{$\boldsymbol{X}^{(i)}$ of Submodel-2}
    \label{alg:submod2}
    \begin{algorithmic}
        \IF{ $u \geq 0$}
        \STATE $\boldsymbol{X}^{(i)} = \boldsymbol{X}^{(1)} $
        \ELSE
        \STATE$\boldsymbol{X}^{(i)} = \boldsymbol{X}^{(2)} $
        \ENDIF
    \end{algorithmic}
\end{algorithm}
Here, \Cref{alg:submod2} shows only $\boldsymbol{X}^{(i)}$, but switches $\boldsymbol{Y}^{(i)},~\boldsymbol{N}^{(i)}$ as well.

VecTwin rudder-equipped vessels do not reverse their propellers but instead gain stopping power by taking a large rudder angle up to 105$\degree$. Therefore, we can add a conditional branch with a nominal stall angle of rudders. The VecTwin rudders have a nominal stall angle of $\alpha_{\text{stall,s}}=50\degree,~\alpha_{\text{stall,p}}=55\degree$ which can be observed from the captive model test results \citep{Watanabe2022vectwin}. Submodel-3 shown in \Cref{alg:submod3} and Submodel-4 shown in \Cref{alg:submod4}. Submodel-3 uses the stall angle as the branch for forwarding conditions only, but Submodel-4 uses it for both forward and astern.

\begin{algorithm}[tb]
    \caption{$\boldsymbol{X}^{(i)}$ of Submodel-3}
    \label{alg:submod3}
    \begin{algorithmic}
        \IF{ $u \geq 0$}
            \IF{$|\delta_s| <\alpha_{\text{stall,s}}~\AND~|\delta_s| <\alpha_{\text{stall,p}}$}
                \STATE $\boldsymbol{X}^{(i)} = \boldsymbol{X}^{(1)}$
            \ELSE
                \STATE $\boldsymbol{X}^{(i)} = \boldsymbol{X}^{(2)}$
            \ENDIF
        \ELSE
        \STATE$\boldsymbol{X}^{(i)} = \boldsymbol{X}^{(3)} $
        \ENDIF
    \end{algorithmic}
\end{algorithm}

\begin{algorithm}[tb]
    \caption{$\boldsymbol{X}^{(i)}$ of Submodel-4}
    \label{alg:submod4}
    \begin{algorithmic}
        \IF{ $u \geq 0$}
            \IF{$|\delta_s| <\alpha_{\text{stall,s}}~\AND~|\delta_s| <\alpha_{\text{stall,p}}$}
                \STATE $\boldsymbol{X}^{(i)} = \boldsymbol{X}^{(1)}$
            \ELSE
                \STATE $\boldsymbol{X}^{(i)} = \boldsymbol{X}^{(2)}$
            \ENDIF
        \ELSE
            \IF{$|\delta_s| <\alpha_{\text{stall,s}}~\AND~|\delta_s| <\alpha_{\text{stall,p}}$}
                \STATE $\boldsymbol{X}^{(i)} = \boldsymbol{X}^{(3)}$
            \ELSE
                \STATE $\boldsymbol{X}^{(i)} = \boldsymbol{X}^{(4)}$
            \ENDIF
        \ENDIF
    \end{algorithmic}
\end{algorithm}

As shown above, the proposed model can apply to various flow fields by adding Submodels because it includes all terms that may be relevant to the characteristics of the flow and actuators. For example, for a single-propeller, single-rudder ship, if the user assumes that characteristics of propeller reversal are not crucial for their usage, then Submodel-2 can be used. Still, if a difference is desired, Submodel-4 can be used.

\subsection{Modeling on wind disturbance}\label{sec:vecwtin_si_wind}
In this study, the wind force is modeled using Fujiwara's regression formula, \citep{Fujiwara1998}, which has fewer model parameters than Isherwood's regression \citep{Isherwood1973} because trigonometric functions represent wind direction. Fujiwara's wind force model is shown by \Cref{eq:fujiwara_model},
    \begin{align}\label{eq:fujiwara_model}
        \begin{aligned}
            X_{A} &= (1/2)\rho_{A}U_{A}^{2}A_{T}\cdot C_{X} \\
            Y_{A} &= (1/2)\rho_{A}U_{A}^{2}A_{L}\cdot C_{Y} \\
            N_{A} &= (1/2)\rho_{A}U_{A}^{2}A_{L}L_{OA}\cdot C_{N} \enspace,
        \end{aligned}
    \end{align}
where
    \begin{align}
        \begin{aligned}
            C_{X} =& X_{A0}+X_{A1} \cos (2\pi - \gamma_{A})+X_{A3} \cos 3 (2\pi - \gamma_{A}) \\
                  &+X_{A5} \cos 5 (2\pi - \gamma_{A}) \\
            C_{Y} =& Y_{A1} \sin (2\pi - \gamma_{A})+Y_{A3} \sin 3 (2\pi - \gamma_{A}) \\
                  &+Y_{A5} \sin 5 (2\pi - \gamma_{A}) \\
            C_{N} =& N_{A1} \sin (2\pi - \gamma_{A})+N_{A2} \sin 2 (2\pi - \gamma_{A}) \\
                  &+N_{A3} \sin 3 (2\pi - \gamma_{A}) \enspace.
        \end{aligned}
    \end{align}
$X_{Ai},~Y_{Ai},~N_{Ai}$ are the model coefficients of wind force. Fujiwara's formula estimates the $X_{Ai},~Y_{Ai},~N_{Ai}$ by regression formulae with the ship's particulars as explanatory variables. But in this study, they were obtained by optimization.

\section{Optimization procedure and its condition}

In this chapter, we describe the method to optimize the parameters of a proposed model using system identification (SI). SI of ship maneuvering is defined as the problem of finding a parameter vector of a dynamic model $\boldsymbol{\theta}\opt$ that minimizes the difference between the state variable history of the input dataset $\mathcal{D}$ and the state variable history estimated by the maneuvering simulation using obtained model. Here, $\mathcal{D}$ is the measured results of model tests or full-scale ship trials and includes the data of state variables $\boldsymbol{s}_{t}$, control input $\boldsymbol{a}_{t}$ and disturbances $\boldsymbol{\omega}_{t}$. The dataset $\mathcal{D}$ consisted of a training dataset $\mathcal{D}\train$ for the optimization procedure, validation dataset $\mathcal{D}\valid$ for selecting the optimal parameter and hyperparameter, and test dataset $\mathcal {D}\test$ to test the generalization performance on unknown data. From now on, bracketed superscripts indicate values for the associated dataset, and superscript (input) indicates one of (train), (validation), or (test).

Hereafter, the chapter will be organized as follows: The objective function, which is the metric for the optimization, is described in \Cref{sec:obj_func}; the domain of the parameter exploration is described in \Cref{sec:range}; the details of the dataset are described in \Cref{sec:dataset}; the hyperparameters are described in \Cref{sec:misc_conditions}; the optimization algorithm CMA-ES in \Cref{sec:cma}; and a model generated by conventional parameter identification method used as a comparison is described in \Cref{sec:efd_model}.

\subsection{Objective function}\label{sec:obj_func}
 The optimization procedure is defined as exploration for the parameter vector $\boldsymbol{\theta}$ of the dynamic model that minimizes the objective function $\mathcal{F}$ for the training data set $\mathcal{D}\train$ using an optimization algorithm, CMA-ES. Objective function $\mathcal{F}$ is defined as follows:
\begin{align}
    \boldsymbol{\theta}\opt\train & =\argmin_{\boldsymbol{\theta} \in \Theta}~\mathcal{F}(\boldsymbol{\theta};~\mathcal{D}\train) \label{eq:thetaopt_vectwin_abko}\\
    \intertext{where}
    \begin{split}
        \mathcal{F}(\boldsymbol{\theta};~\mathcal{D}\train) 
        & = \mathcal{L}\train(\boldsymbol{\theta};~\mathcal{D}\train) \\ 
        &+ \alpha P_{\text{div}}(\boldsymbol{\theta})+ \lambda \|\boldsymbol{\theta}\|_{1} \label{eq:obj_func}\\
    \end{split} \\
    \begin{split}       \mathcal{L}\train&(\boldsymbol{\theta};~\mathcal{D}\train) \\
    &= \sum_{i=1}^{N} \int_{t=0}^{t_{\mathrm{f}}}\| \hat{\boldsymbol{s}_{t}}^{(\text{train},i)}- \hat{\boldsymbol{s}_{t}}^{(\text{sim},i)}(\boldsymbol{\theta})\|^{2} \mathrm{d}t \enspace . \label{eq:norm_def}
   \end{split} 
\end{align} 
The first term of \Cref{eq:obj_func} is the error norm $\mathcal{L}$ of the maneuvering simulation using $\boldsymbol{\theta}$ for the data set $\mathcal{D}\train$. The second term is the deviation penalty from the a priori acceleration range for $\boldsymbol{\theta}$, and $\alpha$ is its weighting coefficient. The third term is the regularization term, and $\lambda$ is the L1 regularization penalty. Detail of the \AddD{first term} is described in \Cref{sec:maneu_sim}, and the second term is described in \Cref{sec:div_penalty}. 
 In \cref{eq:norm_def}, superscript $i=1\ldots N$ denotes the $i$-th contiguous subsequences (CS), and $N$ denotes the total number of CS. Details of CS are stated in the next subsection. Also, $\hat{s}_{t,j}^{(\cdot,i)}$ is the standardized state variables of the $i$-th CS at time $t$, and their $j$-th component is defined by the mean $\mu_{j}^{(\text{train},i )}$ and standard deviation $\sigma_{j}^{(\text{train},i)}$ as in the following equation:
\begin{align}
        &\hat{s}_{t,j}^{(\text{train},i)}=\left(s_{t,j}^{(\text{train},i)}-\mu_{j}^{(\text{train},i)} \right) / \sigma_{j}^{(\text{train},i)} \\
        &\hat{s}_{t,j}^{(\text{sim},i)}=\left(s_{t,j}^{(\text{sim},i)}-\mu_{j}^{(\text{train},i)} \right) / \sigma_{j}^{(\text{train},i)} \enspace.  
\end{align}

\subsubsection{Maneuvering Simulation and Error Norm $\mathcal{L}$}\label{sec:maneu_sim}
The maneuvering simulation is conducted to evaluate the performance of $\boldsymbol{\theta}$, by estimating the state variable $\boldsymbol{s}_{t}\simu$ as an initial value problem using the dynamic model described in \Cref{sec:model} and $\boldsymbol{\theta}$. The control input $\boldsymbol{a}_{t}$ and the disturbance $~\boldsymbol{\omega}_{t}$ are given. In this study, $\boldsymbol{s}_{t_{0}}^{(\text{sim})},~\boldsymbol{a}_{t}^{(\text{sim})},~\boldsymbol{\omega}_{t}^{(\text{sim})}$ were defined by the input dataset $\mathcal{D}^{(\text{input})}$ as \Cref{eq:initial_problem}: 
 \begin{align}\label{eq:initial_problem}
    \begin{aligned}
        \boldsymbol{s}_{t_{0}}^{(\text{sim},i)}&=\boldsymbol{s}_{t_{0}}^{(\text{input},i)} \\ \boldsymbol{a}_{t}^{(\text{sim})}&=\boldsymbol{a}_{t}^{(\text{input},i)} \\ 
        \boldsymbol{\omega}_{t}^{(\text{sim},i)}&=\boldsymbol{\omega}_{t}^{(\text{input},i)} \enspace.
    \end{aligned}
    \end{align}
Here, $\mathcal{D}^{(\text{input})}$ is one of $\mathcal{D}\train,~\mathcal{D}\valid,~\mathcal{D}\test$. We can get the time derivative of $\boldsymbol{s}_{t}$: $\dot{\boldsymbol{s}}_{t}\equiv (\dot{u},~\dot{\vm},~\dot{r})$ by solving \Cref{eq:motion}. The first-order Euler method was used for time development. The time interval $\Delta t=0.1$ s was used.

In the simulation, we divide $\mathcal{D}^{(\text{input})}$ into contiguous subsequences: $\mathcal{D}^{(\text{input},i)}$ to avoid accumulation of errors. We divided $\mathcal{D}^{(\text{input})}$ into CS because the velocity was used in the error norm (\Cref{eq:norm_def}) instead of accelerations. Accelerations can be directly obtained by solving \Cref{eq:motion}; however, 
 acceleration measurement is more difficult than speed and angular velocity measurement. On the other hand, the estimation error of accelerations will be accumulated in the velocity components. The length of the CS is set to $t_{\mathrm{f}}=100$ s.



In addition, because we search for coefficients from a wide range, certain combinations of model coefficients can result in unusually large acceleration during maneuvering simulations. Unrealistically large acceleration will cause numerical overflows in $\boldsymbol{s}_{t}$ or $\mathcal{L}$. We try to prevent overflow by substituting the extra large value with a constant value corresponding to the time-step number when the absolute values of velocity and acceleration exceed the limit $\boldsymbol{S}_{\text{limit}}$, same as \citep{Miyauchi2022SI}: 
\begin{align}\label{eq:xlim_vectwin_abko}
  S_{t,j=1\ldots6}(t) &= 
  \begin{cases}
          \mathrm{sgn}\big(S_{t,j}\big)(2-t/t_{\mathrm{f}})S_{\text{limit},j} & : |S_{t,j}|> S_{\text{limit},j} \\
          S_{t,j} & :\text{else}
      \end{cases}\\
    \intertext{where}
        \boldsymbol{S}_{t} &\equiv (\boldsymbol{s}_{t}, \dot{\boldsymbol{s}}_{t})\\
        \begin{split}           \boldsymbol{S}_{\text{limit}} &=(a,~a,~2a/\lpp,~a,~a,~2a/\lpp)\\  
        a&= 1.0\times10^{80}
        \end{split}\\
        \mathrm{sgn}(x) &= \begin{cases}
        1 & : x > 0 \\
        -1 & :x < 0 \\
        0 & : x = 0
        \end{cases} \enspace.
\end{align}

\subsubsection{Deviation Penalty $P_{\text{div}}$}\label{sec:div_penalty}

Because the proposed model is a simple polynomial equation that does not include hydrodynamic constraints, the acceleration may be overestimated for unknown data, which may cause generalization performance degradation. To prevent overestimation, we imposed a penalty on the objective function when the estimated acceleration exceeded the a priori limit for a specific state and control inputs, which was assumed to give maximum acceleration. This method \citep{Hamada2022} was expected to eliminate model parameters that return unrealistic accelerations.

Below is the detailed calculation method of the deviation penalty, $P_{\text{div}}$. With given initial state variable $\boldsymbol{s}_{0}$, steady control input $\boldsymbol{a}_{t}=\boldsymbol{a}_{0}$, and steady disturbance $\boldsymbol{\omega}_{t}=\boldsymbol{\omega}_{0}$ a penalty is applied if the estimated acceleration $\dot{\boldsymbol{s}}_{t}^{(\text{sim})}$ exceed the predefined range $\left[ \dot{\boldsymbol{s}}_{\text{min}},\dot{\boldsymbol{s}}_{\text{max}} \right]$ as follows:
\begin{align}
  P_{\text{div}} &= \sum_{n=1}^{N_{\text{div}}}P_{n} \\
  \text{where} \nonumber\\
  P_{n}&=\begin{cases}
    1-\frac{ t_\mathrm{over}-\Delta t}{t_{\mathrm{f}}} &: t_\mathrm{over} < t_{\mathrm{f}}^{\text{(div)}}\\
    0 &:\text{else}
  \end{cases}\\
  t_\mathrm{over} &= \inf \left\{ t \geq 0 : \boldsymbol{\dot{s}}_{t}^{(\text{sim})} \not \in \left[ \dot{\boldsymbol{s}}_{\text{min}},\dot{\boldsymbol{s}}_{\text{max}} \right] \right\} \enspace.
\end{align}
Here $t_{\mathrm{f}}^{\text{(div)}}$ is duration of the maneuvering simulation and $t_{\mathrm{f}}=1$ s. Since  $\boldsymbol{s}_{0}$ and $\boldsymbol{a}_{0}$ are  expected combinations to generate the maximum or minimum acceleration, we set $s_{0}=(u_{0},~v_{0},~r_{0})$ by the maximum and minimum value of $\mathcal{D}$ with a certain margin, and set $a_{0}=(n_{\mathrm{P}0},~\delta_{s0},~\delta_{p0})$ by the value which expected to generate maximum and minimum control forces. Accordingly, $\boldsymbol{s}_{0}$ and $\boldsymbol{a}_{0}$ are show as in \Cref{eq:aug_dataset}:
\begin{align}\label{eq:aug_dataset}
  \begin{aligned}
    u_{0} &\in \{-0.2, 0.0, 0.5\} \\
    v_{0} &\in \{-0.2, 0.0, 0.2\} \\
    r_{0}     &\in \{-0.1, 0.0, 0.1\} \\
    n_{\mathrm{P}0} &\in \{0.0, 12.5\} \\
    \delta_{s0} &\in \{-35, 0, 45, 60, 90\} \\
    \delta_{p0} &\in \{35, 0, -45, -60, -90\} \enspace .
  \end{aligned}
\end{align}
Total number of combination $N_{\text{div}}$ is 1350. In addition, no wind disturbance is assumed $\boldsymbol{\omega}_{0}=(0,~0)$.

The range of $\left[\dot{\boldsymbol{s}}_{\text{min}},\dot{\boldsymbol{s}}_{\text{max}} \right]$ was defined using a priori information. Here, we used 1-second moving average of the numerical differentiation of $\boldsymbol{s}_{t}^{(\mathcal{D})}$ as the acceleration $\dot{\boldsymbol{s}}_{t}^{(\mathcal{D})}$. This study's model experiment did not measure acceleration due to its low signal-noise ratio. The 99.7th percentile of the of modulus of the acceleration $\boldsymbol{Q}_{99.7}(|\dot{\boldsymbol{s}}_{t}^{(\mathcal{D})}|)$ was used to set the range:
\begin{equation}\label{eq:set_sdot}
    \left[ \dot{\boldsymbol{s}}_{\text{min}},\dot{\boldsymbol{s}}_{\text{max}} \right] = [-\boldsymbol{Q}_{99.7}(|\dot{\boldsymbol{s}}_{t}^{(\mathcal{D})}|),~\boldsymbol{Q}_{99.7}(|\dot{\boldsymbol{s}}_{t}^{(\mathcal{D})}|)]\enspace .
\end{equation}
%
We can simplify the range by employing the modulus function in \Cref{eq:set_sdot}. The reason for using the 99.7th percentile is to remove the effect of noise amplified by numerical differentiation. If the subject ship's acceleration can be measured appropriately, measured acceleration is more suitable.

\subsection{Exploration domain}\label{sec:range}
In this section, we show the exploration domain $\boldsymbol{\Theta}$ of parameter $\boldsymbol{\theta}$. The parameter vector $\boldsymbol{\theta}$ is consisted of added masses and inertia, hydrodynamic coefficients, and wind force coefficients, as follows:
\begin{equation}
    \begin{split}
        \boldsymbol{\theta} = 
        & \big(m_{x},~m_{y},~I_{zz}+J_{zz},~ \\
        &\boldsymbol{X}^{(i)},~\boldsymbol{Y}^{(i)},~\boldsymbol{N}^{(i)},\\
        &X_{A0},~X_{A1},~X_{A3},~X_{A5}, \\ 
        &Y_{A1},~Y_{A3},~Y_{A5},~
        N_{A1},~N_{A2},~X_{A3}~\big) \enspace .
    \end{split}
\end{equation}


Domain $\boldsymbol{\Theta}$ must be defined without a captive model test of the subject ship a priori. In our previous study \citep{Miyauchi2022SI}, boundary of the domain $\boldsymbol{\Theta}$ is set to be approximately 10 times greater value of a priori solutions, i.e., captive model test results and empirical value. Because of this vast exploration domain, we believed we could obtain optimal coefficients even if the hull form or dynamic model's formulae differed from a priori solutions. Hence, we set the domain boundary by multiplying the coefficient values of the classic Abkowitz model for single-propeller, single-rudder ship \citep{Abkowitz1980}:
\begin{equation}\label{eq:rangeall_vectwin_abko}
    \theta_{j} \in \Theta_{j} =\big[-1.0, 1.0\big] \enspace.
\end{equation}
Here, $\Theta_{j}$ is a domain for the $j$-th component of the parameter vector.
Moreover, several exceptions of domain boundary setting were made, like the previous study, as follows:
\begin{itemize}
\item Maximum value of $X_{u},~X_{uu},~Y_{\vm},~N_{r}$ are set to 0 because the signs of diagonal resistance components are self-explanatory.
\item Minimum value of $X_{\np}$ was set to 0 because the sign of the first-order thrust coefficient is self-explanatory.
\item We multiplied $10$ to \Cref{eq:rangeall_vectwin_abko} according to the order of $r$ contained in $\theta_{j}$, because $r$ is relatively smaller than the other components of $\boldsymbol{x}$ due to its radian/s dimension.
\item We multiplied $0.1$ to \Cref{eq:rangeall_vectwin_abko} according to the order of $\np$ contained in $\theta_{j}$, because the subject ship is a model ship, $\np$ is relatively larger than the other components of $\boldsymbol{x}$.
\item The added mass, added mass moment, and wind force coefficients were determined from the empirical value $\theta_{\text{EFD},j}$. For $m_{x},~m_{y},~I_{zz}+J_{zz}$, $\Theta_{j}=\big[-10 \theta_{\text{EFD},j},~ 10\theta_{\text{EFD},j}\big]$ and for wind coefficient, $\Theta_{j}=\big[-10 \theta_{\text{EFD},j},~ 10\theta_{\text{EFD},j}\big]$.
\end{itemize}

There are Pros and Cons to searching dimensional values. In dimensional form, domain $\boldsymbol{\Theta}$ depends on the ship's size; hence,$\boldsymbol{\Theta}$ needs to be determined by repeated trials every time the subject ship changes. 
In contrast, $\boldsymbol{\Theta}$ for non-dimensional coefficients can set its domain boundaries independent of the size and scale of the ship; however, non-dimensionalization causes a problem when ship speed is zero. Further study on domain design is needed.

\subsection{Dataset}\label{sec:dataset}
This section describes the datasets used in the study. Maneuvering time histories of a 3-meter free-running model ship were used as the datasets. The measurement system of the model ship is almost the same as in the previous study \citep{Miyauchi2022SI}, with the following differences. Ship speed $u$ and $\vm$ were calculated from Speed Over Ground (SOG), Course Over Ground (COG) measured by the GNSS receiver, and ship heading measured by fiber optical gyro (FOG).  In addition, compared to previous studies, the current FOG has better heading accuracy; the yaw drift was about $1\degree$/hour. Moreover, unlike in previous studies, $r$ was not filtered due to the improved performance of FOG.

Random maneuvers are the maneuver used in the dataset, same as the previous study \citep{Miyauchi2022SI}. The reason for using random maneuvers was used is to ensure the stability of the dynamic model (\Cref{req:stability}), as described in \Cref{sec:model}. The random maneuvers are intended to obtain various different values of state variables and control inputs that can occur as the motion of the subject ship as possible. In this study, as in the previous studies, the model ship operator gives random $\boldsymbol{a}_{t}$ so that the state variables and control inputs are distributed. For details on the random maneuvers, please refer to the previous study \citep{Miyauchi2022SI}. The experiment was conducted in an experiment  pond (\textit{Inukai} Pond) at Osaka University.

The control input $\boldsymbol{a}_{t}$ of random maneuvers was given within a set upper and lower limit range, as shown in \Cref{tab:control range}. The range of $\deltas,~\deltap$ is the maximum mechanical steering range of the VecTwin rudder. With the maximum propeller speed $n_{\mathrm{p}\max}$, the ship reaches $u\approx0.6$ m/s in a forwarding equilibrium state. The dimensionless speed corresponding to $u=0.6$ is $\mathrm{Fr}=u/\sqrt{g\lpp}=0.11$, equivalent to 6.7 knots for a medium-sized ship (assuming $\lpp=100$ m). Therefore, $n_{\mathrm{p}\max}$ is a typical range of propeller speed for ship handling in a harbor. Since the subject ship is equipped with VecTwin rudder system, the ship can stop and astern without reversing the propeller; hence, $n_{\mathrm{p}\min}=0$ was used.

\begin{figure*}
    \centering
    \includegraphics[width=\linewidth]{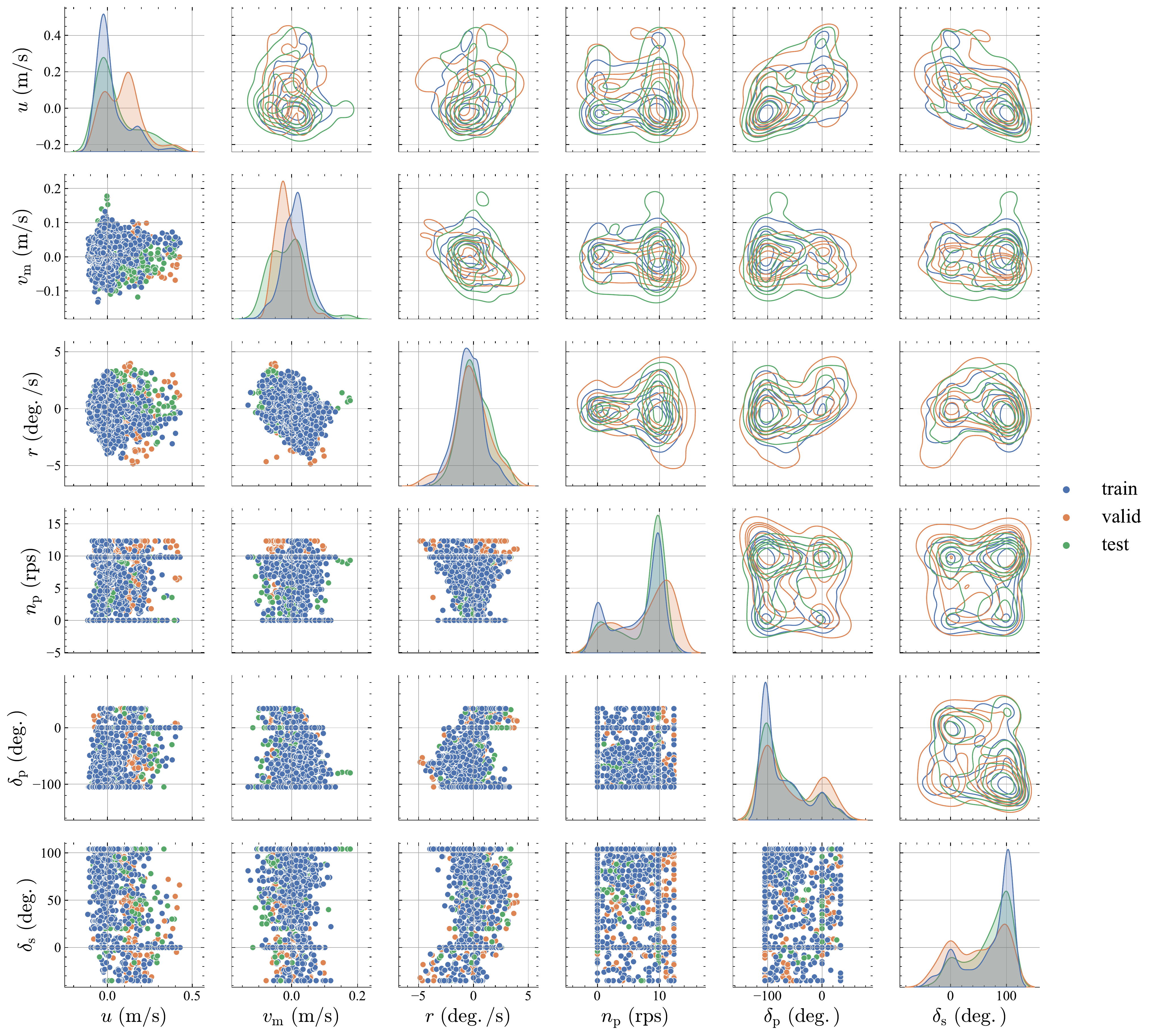}
    \caption{Distribution of state variables and control inputs of data sets. Diagonal and upper triangle figures are univariate and bivariate kernel density estimations of state variables and control inputs, respectively. Lower triangle figures are bivariable scatter plots of state variables and control inputs. In this figure, data sets were re-sampled with one sample per five seconds for plotting.}
    \label{fig:data_set_dist}
\end{figure*}

The entire dataset $\mathcal{D}$ was divided into $\mathcal{D}\train,~\mathcal{D}\valid$ and $\mathcal{D}\test$, and here we compare the distribution among the datasets. First, the number of samples for each dataset is shown in time in \cref{tab:duration}. The measurement frequency of the datasets is $0.1$ s, equal to $\Delta t$ of the maneuvering simulation. The entire dataset $\mathcal{D}$ is divided to be the length of each dataset $T\train,~T\valid,~T\test$ is approximately $6:1:1$. The distribution of $\boldsymbol{s}_{t}$ and control inputs $\np,~\deltas,\deltap$ for each dataset are shown in \Cref{fig:data_set_dist}. The distributions of the datasets generally agree well with each other.

The data set $\mathcal{D}$ is expected to cover all possible maneuvers that may occur during harbor maneuvers, including unknown maneuvers, to obtain a robust dynamic model. There is no clear definition of the range of state variables for harbor maneuvers; however, the state variables in $\mathcal{D}$ are comparable to the empirical range of speeds used in the harbor when converted to real ship scale. Specifically, $u$ is $u \in [-0.1,~0.4]$, equivalent to 1.2 knots astern and 4.5 knots forward of $\lpp=100$. Both $\vm$ and $r$ are approximately symmetric, i.e., positively and negatively symmetrically distributed, and $\vm \in [-0.1,~0.1]$, which is the range of slow maneuver corresponding to $\mathrm{Fr}_{\vm}=\pm 0.018$ and 1.2 knots at $\lpp=100$. Further, by observing the distribution of $u$ and $\vm$, we see that the drift angle $\beta$ is distributed over the entire circumference.

In addition, random maneuvers are favorable because correlations between state variables and control inputs are weak. The multi-colinearity is a known problem when using Zig-zag maneuvers and turnings as training data due to the strong correlation between $\vm$ and $r$ in those motions \citep{Abkowitz1980,Hwang1982}. On the other hand, in the random maneuvers, the correlation between all state variables and control inputs is weak, not only between $\vm$ and $r$ as shown in \Cref{fig:data_set_dist}.

In this study, random maneuvers were employed as the data set, but the validity of the proposed model needs to be demonstrated for motions other than random maneuvers. Here, other than random maneuvers, Crash-Astern (CA) test data set $\mathcal{D}\ca$ was employed. A CA test was performed by making a steady propeller revolution at $\np=7.3$ rps or $\np=9.8$ rps from a stopped state, accelerating straight ahead at $\delta_{\mathrm{sp}}=0\degree$ for 120 s ($\np=7.3$) or 100 s ($\np=9.8$), and steer $\delta_{\mathrm{s}}=105\degree,~\delta_{\mathrm{p}}=-105\degree$ to decelerate, stop, and astern. In total, $\mathcal{D}\ca$ contains four CA tests, and each test is about 400 seconds. Hence the total length of the $\mathcal{D}\ca$ is $T\ca=1647.7$ s as shown in \Cref{tab:duration}.

\begin{figure}
    \centering
    \includegraphics[width=\linewidth]{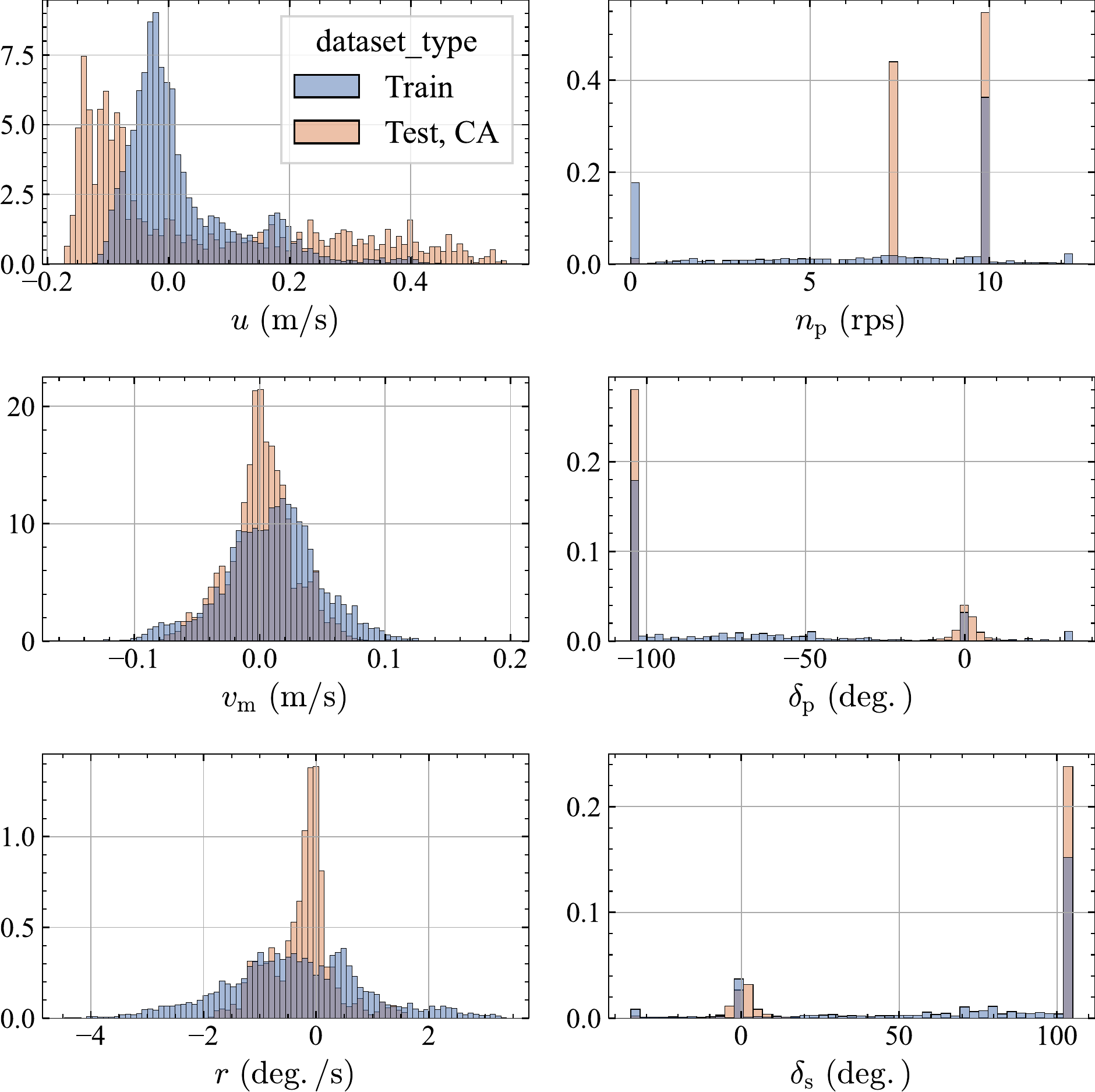}
    \caption{Histograms of $x_{k}$ of $\mathcal{D}\train$ and $\mathcal{D}\ca$. Histograms are normalized to make the total area of the bins to be 1 for each data set.}
    \label{fig:hist_train_vs_ca}
\end{figure}

The normalized histogram of $\mathcal{D}\ca$ and $\mathcal{D}\train$ shown in \cref{fig:hist_train_vs_ca}. As mentioned earlier, CA test only used a specific $\np,~\deltas,~\deltap$ as a command signal, which results in a stronger $\np,~\deltas,~\deltap$ bias than the random maneuvers. In addition, since $\np,~\deltas,~\deltap$ were constant, the speed developed faster, and the distribution of $u$ was wider than that of random maneuvering. On the other hand, $\vm$ and $r$ were more biased around 0 than in random maneuvers.

\begin{table}
    \centering
    \caption{Duration of datasets.}
    \begin{tabular}{ccccc}
    \toprule
        Name  & $T\train$ & $T\valid$ & $T\test$ & $T\ca$\\
        \midrule
        Value (s)& $7407.4$ & $1201.2$ & $1201.2$ & $1647.7$ \\
        \bottomrule
    \end{tabular}
    \label{tab:duration}
\end{table}

\begin{table}
    \centering
    \caption{Maximum and minimum of control input $\np,~\deltas,~\deltap$ of random maneuver data set $\mathcal{D}$.
        \label{tab:control range}}
    \begin{tabular}{cc}
    \toprule
        Parameter & Maximum and minimum \\
        \midrule
        $\np$ (rps)& $[0,~ 12.5]$ \\
        $\deltas$ (deg.) & $[-35, ~105]$ \\
        $\deltap$ (deg.) & $[-105, ~35]$ \\
        \bottomrule
    \end{tabular}
\end{table}

\subsection{Numerical Conditions of Dynamic Models and hyperparameters}\label{sec:misc_conditions}
The objective function of \Cref{eq:obj_func} requires the user to select the hyperparameters $\alpha$ and $\lambda$. The $\alpha$ and $\lambda$ depend on the dynamic model's complexity, i.e., order of the model's polynomials and number of Submodels. In this study, order of polynomials and the number of Submodels were calculated for five combinations as shown in  \Cref{tab:model_setting}. Hereafter, a dynamic model with Submodel-2 and 2nd-order polynomials will be referred to as a ``SM-2, 2nd-order model.'' In addition, hyperparameters $\alpha$ and $\lambda$ are used for the combinations shown in\Cref{tab:hypparam_2nd,tab:hypparam_3rd}. Hereafter, an optimization performed with a specific dynamic model, $\alpha$, and $\lambda$ is referred to as a ``computational case.'' A total of 28 computational cases were executed.
\begin{table}
    \centering
    \caption{Maximum order of polynomials and number of Submodels of the dynamic models.}
    \begin{tabular}{ccc}
    \toprule
        Submodels & Max. order  & Unknown variables \\
         \midrule
        2 &  \multirow{3}{*}{2nd order} & 133 \\
        3 &  & 193 \\
        4 &  & 253 \\
        \midrule
        2 & \multirow{2}{*}{3rd order} &  413 \\
        3 & & 613 \\
        \bottomrule
    \end{tabular}
    \label{tab:model_setting}
\end{table}



\begin{table}
    \centering
    \caption{Settings of computational cases with 2nd order models.}
    \begin{tabular}{ccccc}
        \toprule
        Case No. & Order & SM & $\alpha$ & $\lambda$ \\
        \midrule
        1 & \multirow{4}{*}{2nd} & \multirow{4}{*}{2} & 0 &0 \\
        2 &  &  & 0     & $1e2$ \\
        3 &  &  & $1e2$ &0 \\
        4 &  &  & $1e2$ & $1e2$ \\
        \midrule
        5 &  \multirow{4}{*}{2nd}& \multirow{4}{*}{3} & 0 &0 \\
        6 &  &  & 0     & $1e2$ \\
        7 &  &  & $1e2$ &0 \\
        8 &  &  & $1e2$ & $1e2$ \\ 
        \midrule
        9 &  \multirow{4}{*}{2nd}& \multirow{4}{*}{4} & 0 &0 \\
        10 &  &  & 0     & $1e2$ \\
        11 &  &  & $1e2$ &0 \\
        12 &  &  & $1e2$ & $1e2$ \\
        \bottomrule
        \end{tabular}
            \label{tab:hypparam_2nd}
\end{table}
\begin{table}
    \centering
    \caption{Settings of computational cases with 3rd order models.}
    \begin{tabular}{ccccc}
        \toprule
        Case No. & Order & SM & $\alpha$ & $\lambda$ \\
        \midrule
        13 & \multirow{8}{*}{3rd} & \multirow{8}{*}{2} & 0 &0 \\
        14 &  &  & 0     & $1e2$ \\
        15 &  &  & 0     & $1e4$ \\
        16 &  &  & $1e2$ &0 \\
        17 &  &  & $1e2$ & $1e2$ \\
        18 &  &  & $1e2$ & $1e4$ \\
        19 &  &  & $1e4$ &0 \\
        20 &  &  & $1e4$ & $1e2$ \\
        \midrule
        21 & \multirow{8}{*}{3rd} & \multirow{8}{*}{3} & 0 &0 \\
        22 &  &  & 0     & $1e2$ \\
        23 &  &  & 0     & $1e4$ \\
        24 &  &  & $1e2$ &0 \\
        25 &  &  & $1e2$ & $1e2$ \\
        26 &  &  & $1e2$ & $1e4$ \\
        27 &  &  & $1e4$ &0 \\
        28 &  &  & $1e4$ & $1e2$ \\
        \bottomrule
        \end{tabular}
   \label{tab:hypparam_3rd}
\end{table}
\subsection{CMA-ES and its conditions}\label{sec:cma}



This study used the optimization method Covariance Matrix Adaption-Evolution Strategy (CMA-ES) \citep{Hansen2006}. CMA-ES is effective for complex optimization problems where the problems are non-separable and multi-modal \citep{10.1145/1830761.1830790,rios2013derivative}. In addition, CMA-ES has strong search capability for optimization problems up to several hundred dimensions\citep{10.1162/evco_a_00260}. We expect CMA-ES to be effective for the Black-Box optimization problem we are tackling in this study. In addition, CMA-ES has a practical advantage in requiring fewer hyperparameters to be determined by the user.

CMA-ES uses the normal distribution to generate candidate solutions and updates the statistics (mean and covariance matrices) of the normal distribution using candidates with a high degree of adaption to the objective function among the candidates. This process of generation, evaluation, and update is iterated, which is interpreted as minimizing the expectation value of the evaluation function \citep{akimoto2012theoretical,JMLR:v18:14-467}. 

In this study, the CMA-ES with box-constraints\citep{Sakamoto2017} and the restart strategy \citep{Auger2005} was used, as in the previous study \citep{Miyauchi2022SI,Miyauchi2021Abkowitz}. The initial and maximum population sizes of candidate solutions were set to 64 and 256, respectively. This study's optimization using CMA-ES was terminated at $5\times10^5$ iterations.

 \subsection{MMG-EFD model}\label{sec:efd_model}

The proposed model needs to be verified whether it can be used as a maneuvering estimation module of a harbor maneuvering simulator. The authors confirmed the validity of the proposed method by confirming that the proposed model has the same or better estimation performance as the model identified by conventional methods. That is, the low-speed maneuvering MMG model and captive model tests. Hereafter, the model for comparison will be referred to as the MMG-EFD (Experimental Fluid Dynamics) model.

Details of the model and parameters used in MMG-EFD model are as follows. The MMG model estimates the hydrodynamic forces generated by each module, such as the hull, propeller, rudder, and wind force. The sub-modules used in MMG-EFD model are as follows: Yoshimura's model \citep{Yoshimura2009a} for the hull; Fujiwara's model \citep{Fujiwara1998} for the wind pressure; and Kang's model \citep{Kang2008} for the propeller and rudder forces. Model parameters were obtained by a captive model test of the subject ship, empirical formulae, or substituted by other ship's parameters. Specifically, for the hull model, the linear coefficient and the astern drag coefficient were derived from captive model tests, while the other values were obtained by empirical formulae \citep{Yoshimura2009a}. For the propeller and rudder models, propeller thrust coefficients $K_{T0},~K_{T1},K_{T2}$ were obtained from the propeller open test of the subject ship's propeller, and the other coefficients were substituted by VLCC's values given by Kang \citep{Kang2008}. Wind coefficients were derived by Fujiwara's regression formulae using the subject ship's geometric parameters. Although the MMG-EFD model used surrogate parameters for more than half of its parameters, the authors believe that the MMG-EFD model has sufficient accuracy for practical use because the authors used the MMG-EFD model to train an automatic berthing controller and successfully demonstrated a scaled-model experiment of automatic berthing \citep{Wakita2022}.

\section{Results}\label{sec:results}
In this chapter, we examine the performance of the optimal parameter $\boldsymbol{\theta}\opt$ obtained from the optimization. Firstly, we select the optimal parameter for each computational case by observing the learning curve of train loss $\mathcal{L}(\boldsymbol{\theta};~\mathcal{D}\train)$ and validation loss $\mathcal{L}(\boldsymbol{\theta};~\mathcal{D}\valid)$. Secondly, since CMA-ES is a stochastic search method, we check the dependency on the random seed used in CMA-ES. Thirdly, we select the optimal value of hyperparameters by comparing the $\mathcal{L}(\boldsymbol{\theta}_{\text{opt}};~\mathcal{D}\valid)$ with different hyperparameter values.
Finally, the generalization performance is checked using random maneuvering test set $\mathcal{D}\test$ and crash-astern test set $\mathcal{D}\ca$. We also check how the performance varies with the order of model formulae and the Submodel. Furthermore, the estimation error in Train, Validation, and Test are discussed.

The error norm $\mathcal{L}$ defined by \Cref{eq:norm_def} is used as a performance metric. This is because the objective function $\mathcal{F}$ shown in \Cref{eq:obj_func} is penalized by two terms other than the error norm. 

\subsection{Learning Curve of Optimization}
\begin{figure}
            \centering
        \begin{minipage}[c]{\linewidth}
            \centering
            \includegraphics[keepaspectratio, width=\columnwidth]{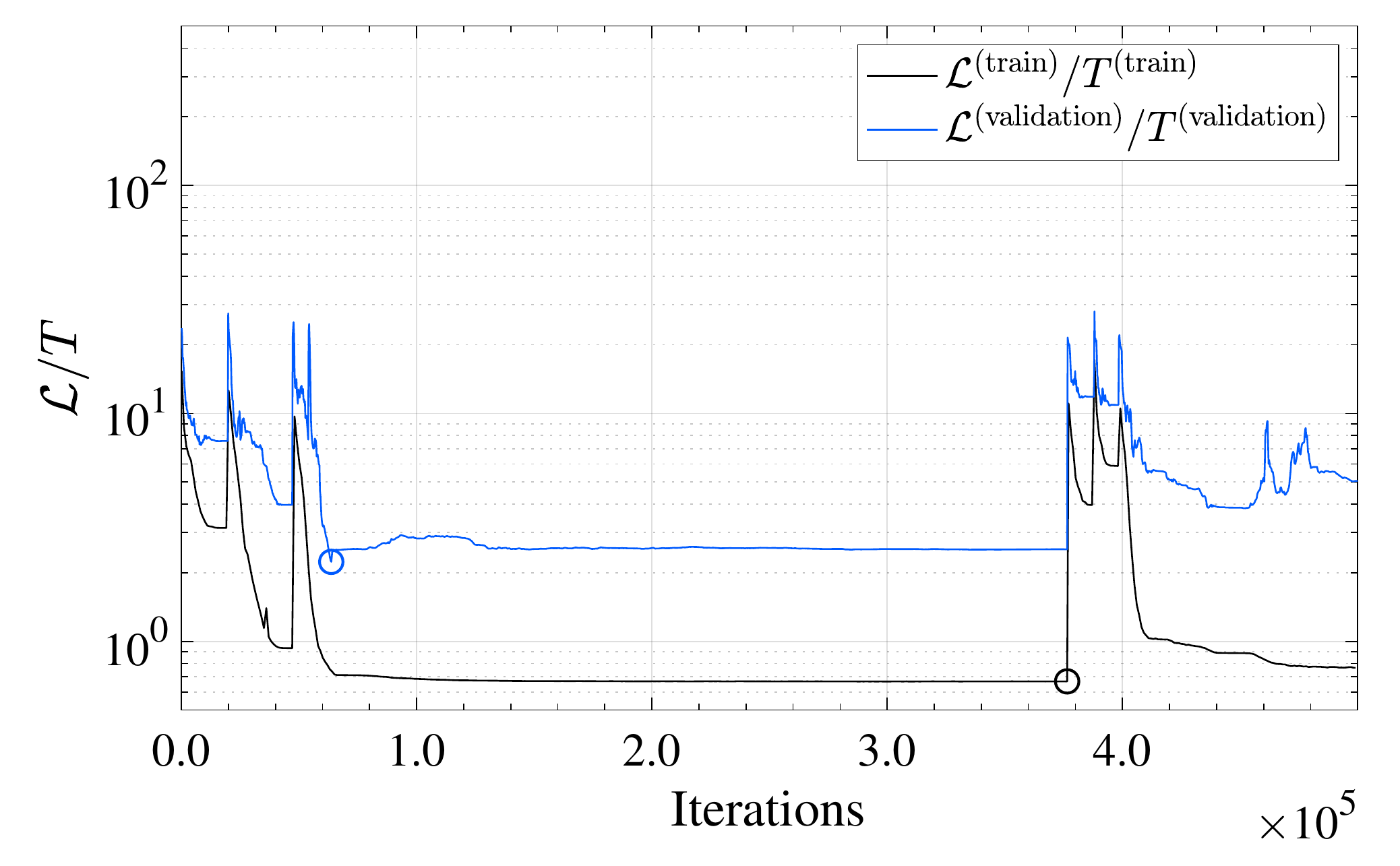}
            \subcaption{SM-2, 3rd order model, $\alpha=0,~\lambda=0$.}
            \label{fig:iter_sm2-3}
        \end{minipage}
        \begin{minipage}[c]{\linewidth}
            \centering
            \includegraphics[keepaspectratio, width=\columnwidth]{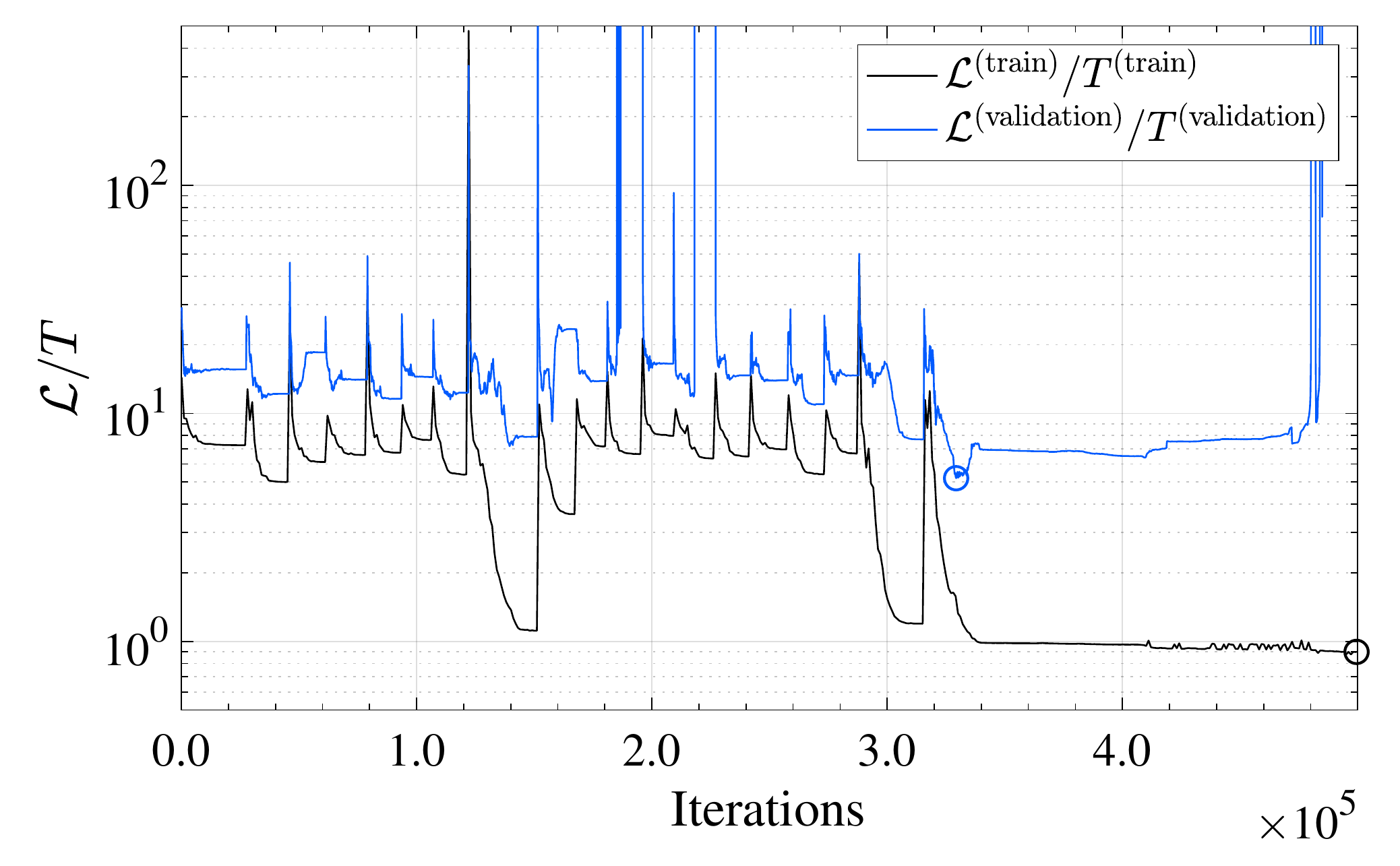}
            \subcaption{SM-3, 3rd order model, $\alpha=0,~\lambda=0$. }
        \label{fig:iter_sm3-3}
        \end{minipage} 
    \caption{Train loss and validation loss through the optimization by CMA-ES. Black circle represents $\mathcal{L}(\boldsymbol{\theta}\train\opt;\mathcal{D}\train)$ and blue circle represents $\mathcal{L}(\boldsymbol{\theta}\valid\opt;\mathcal{D}\valid)$. Note that in \cref{fig:iter_sm3-3}, line plots of  $\mathcal{L}\valid/T\valid$ are not connected when the error norm is much greater than max of vertical axis range: $\mathcal{L}\valid/T\valid \gg 5\times 10^{2}$. Other peaks in the figure are due to the restart strategy of CMA-ES. "SM" stands for Submodel.}
    \label{fig:iter_process}
\end{figure}

The optimization yields the optimal parameter $\boldsymbol{\theta}\opt\train$ in $\mathcal{D}\train$ defined by \cref{eq:thetaopt_vectwin_abko}, but overfitting may occur. Thus, the error norm $\mathcal{L}\valid\equiv \mathcal{L}(\boldsymbol{\theta};~\mathcal{D}\valid)$ on $\mathcal{D}\valid$ shown below,
\begin{align}
    \begin{split}    
    \mathcal{L}&(\boldsymbol{\theta};~\mathcal{D}\valid) \\
    &= \sum_{i=1}^{N} \int_{t=0}^{t_{\mathrm{f}}}\| \hat{\boldsymbol{s}_{t}}^{(\text{validation},i)}- \hat{\boldsymbol{s}_{t}}^{(\text{sim},i)}(\boldsymbol{\theta})\|^{2} \mathrm{d}t \end{split}\label{eq:valid_norm_def}
\end{align}
and $\mathcal{L}\train$ are observed during iterative process of optimization. \Cref{fig:iter_process} shows the learning curve of two representative computational cases. Note that in \Cref{fig:iter_process}, the mean of the candidate solutions $\overline{\boldsymbol{\theta}}$ of each iteration of the CMA-ES were used for error norm $\mathcal{L}\train,~\mathcal{L}\valid$ by \Cref{eq:norm_def,eq:valid_norm_def}. Since the total time of datasets is different, the error norms were compared in error-per-time from ($\mathcal{L}/T$) using the data set time $T\train,~T\valid$. Note that $\mathcal{L}\train$ and $\mathcal{L}\valid$ have several peaks at the same iterations in \cref{fig:iter_process}. This is due to the restart strategy of CMA-ES.



The learning curve of the SM-2, 3rd order model is shown in \Cref{fig:iter_sm2-3}. The validation loss $\mathcal{L}\valid$, indicated by the blue line, showed the same downward trend as training loss $\mathcal{L}\train$. The validation loss stagnated at the value with slight degradation from its minimum indicated by the blue circle. Therefore, we concluded that overfitting did not occur in this computational case. On the other hand, in the SM-3, 3rd order model, shown in \Cref{fig:iter_sm3-3},  $\mathcal{L}\valid$ worsened in the iteration around $\bar{\boldsymbol{\theta}}\opt\train$ indicated by the black circle; thus, we concluded that overfitting has occurred.

To avoid overfitting, the optimal parameter on $\mathcal{D}\valid$ was selected as the optimal parameter for each computational case, and the dynamic model with the optimal parameter $\boldsymbol{\theta}\opt\valid$ was defined as the optimal model. More precisely, the optimal parameter $\boldsymbol{\theta}\opt\valid$ was selected from the set $\Theta\train$, which is the set of  $\overline{\boldsymbol{\theta}}$ per 200 iterations:
\begin{align}
    \boldsymbol{\theta}\opt\valid & =\argmin_{\overline{\boldsymbol{\theta}} \in \Theta\train}~\mathcal{L}(\overline{\boldsymbol{\theta}};~\mathcal{D}\valid) \enspace .\label{eq:validopt}
\end{align}

\subsection{Random seed trial}\label{sec:random_seed}

Since CMA-ES is a stochastic search method, the optimal parameter may depend on the random seed. Before testing the performance of the optimal model, we check the effect of the random seed. Here, we performed five independent trials with different random number seeds for representative cases. Random seed trials with regularization penalty $\lambda$ and deviation penalty $\alpha$ were only conducted for 3rd-order models. Trials on the cases with $\alpha=0$ or $\lambda = 0$ were conducted both on 2nd-order and 3rd-order models.
From now on, for simplicity of notation, the error norm is expressed in the following expression: 
\begin{equation}\label{eq:notation_train_valid}
    \begin{aligned}
      \Ltrainprime &=\mathcal{L}(\boldsymbol{\theta}\opt\train; \mathcal{D}\train)/T\train \\
      \Lvalidprime &=\mathcal{L}(\boldsymbol{\theta}\opt\valid; \mathcal{D}\valid)/T\valid \\
    \end{aligned} \enspace.
\end{equation}

\begin{figure}
    \centering
    \includegraphics[width=\linewidth]{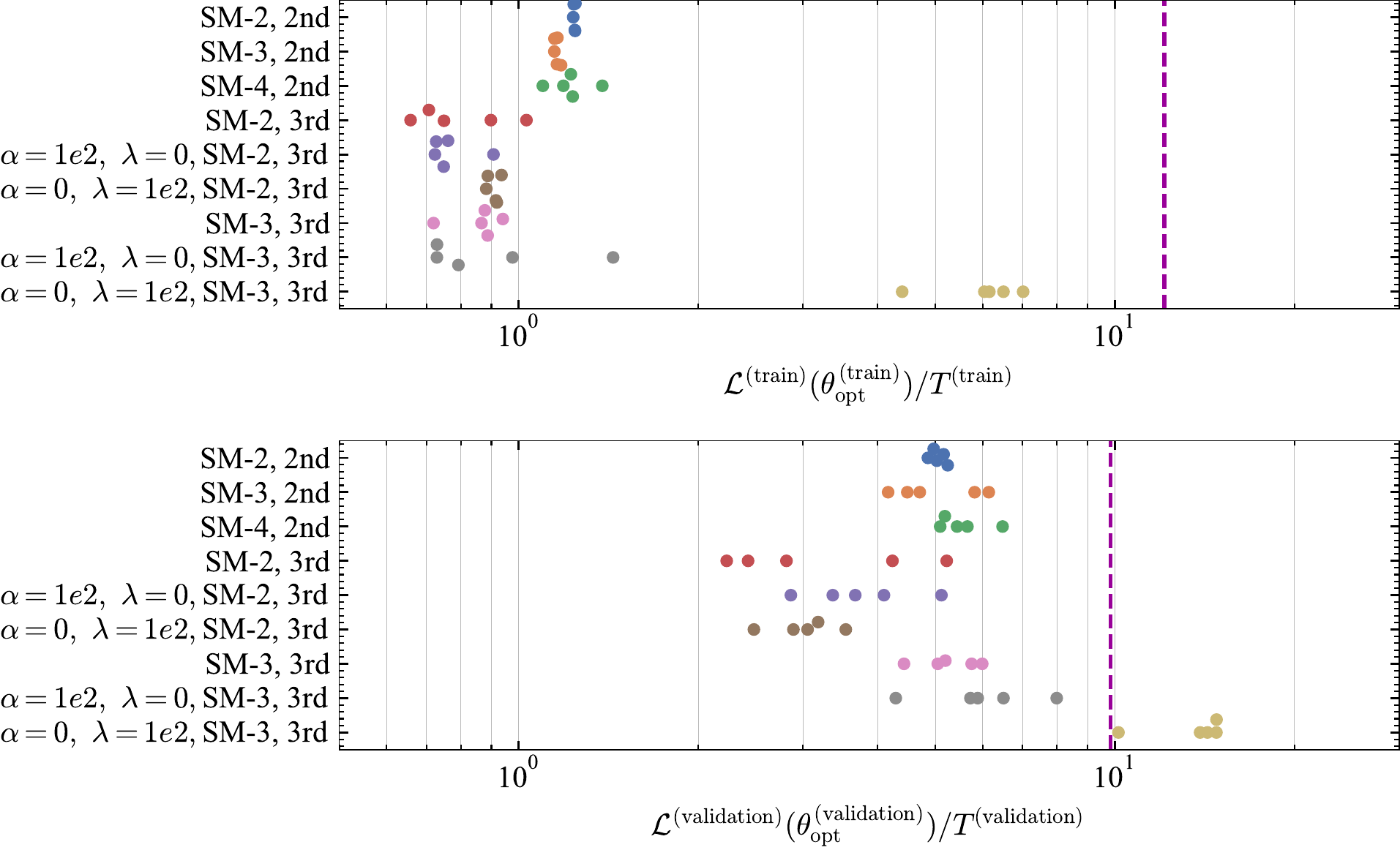}
    \caption{Results of random seed trial. Dot plots represent each random seed trial.  Labels on vertical axis without $\alpha$ and $\lambda$ values, those are $\alpha=0,~\lambda=0$. The vertical purple dashed line represents MMG-EFD model results.} 
    \label{fig:randomseed}
\end{figure}

The effect of random seed on the error norm $\mathcal{L}$ is shown in \cref{fig:randomseed}.
In all cases except for the SM-3, 3rd order model, $\alpha=0,~\lambda=1e2$ shown in yellow, the error norm is smaller than that of the MMG-EFD model for both training and validation dataset regardless of the random seed, indicating that the dynamic model obtained by the proposed method has the same or better estimation performance than the existing methods. 

Next, we discuss the differences in performance for each computational case on the training and validation datasets. First, we focus on $\Ltrainprime$. In $\Ltrainprime$, the third-order model has a smaller error norm $\Ltrainprime$ than the second-order model, which means that it fits the training data better. The effect of the regularization penalty $\lambda$ and the deviation penalty $\alpha$ differed depending on the dynamic model. In the SM-2, 3rd order model, $\alpha,~\lambda$ tended to suppress the variation of $\Ltrainprime$. On the other hand, in the SM-3, 3rd order model, the variation of $\Ltrainprime$ was larger. Moreover, on $\lambda=1e2$ case, $\Ltrainprime$ was significantly worsen. Therefore, the effect of $\alpha$ and $\lambda$ on optimization is different for each dynamic model. A parameter study of the hyperparameters is required for each model when using the proposed method.

Next, we focus on $\Lvalidprime$. Although $\Lvalidprime$ is several times worse than that of $\Ltrainprime$, $\Lvalidprime$ is still smaller than that of the MMG-EFD model except for one case. In other words, the proposed method has the same or better estimation accuracy than the conventional method even when the data is not used for training, as long as the appropriate $\alpha$ and $\lambda$ were selected, regardless of the random seed.

The performance rank of the dynamic models on $\mathcal{D}\valid$ is the same as for training: SM-2, 3rd order model is the best, followed by SM-3, 3rd order model. The second-order model is inferior to the third-order model, and the performance difference between the second-order models is insignificant compared to the variation due to random seeding. Moreover, as in training, the trend of the regularization penalty $\lambda$ and the deviation penalty $\alpha$ differed among the dynamic models. In the SM-2, 3rd order model, the regularization penalty $\lambda$ and the deviation penalty $\alpha$ suppressed the variation of $\Lvalidprime$. In SM-3, 3rd order model, $\alpha=0,~\lambda =1e2$ is an exception where $\Lvalidprime$ is worse than the other cases.
This may be because proper learning is not achieved as described in the previous paragraph, resulting in worse performance for $\Lvalidprime$.

\subsection{Select the Best hyperparameter}\label{sec:train_vs_valid}
\begin{figure}
    \centering
    \begin{minipage}[b]{\linewidth}
            \includegraphics[keepaspectratio, width=\columnwidth]{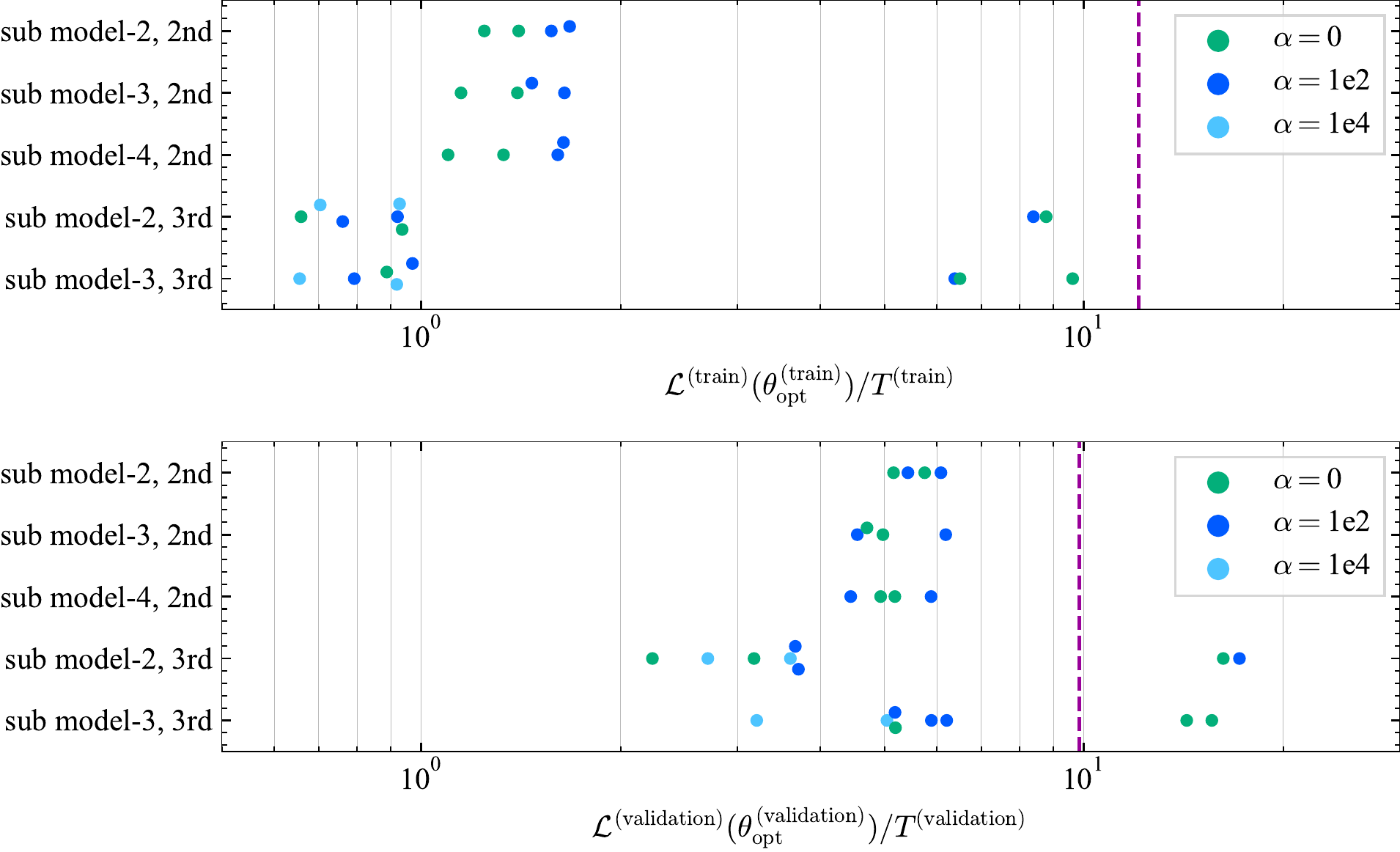}
            \subcaption{Color labeled by value of $\alpha$.}
            \label{fig:alpha_results}
    \end{minipage}\\
    \begin{minipage}[b]{\linewidth}
            \includegraphics[keepaspectratio, width=\columnwidth]{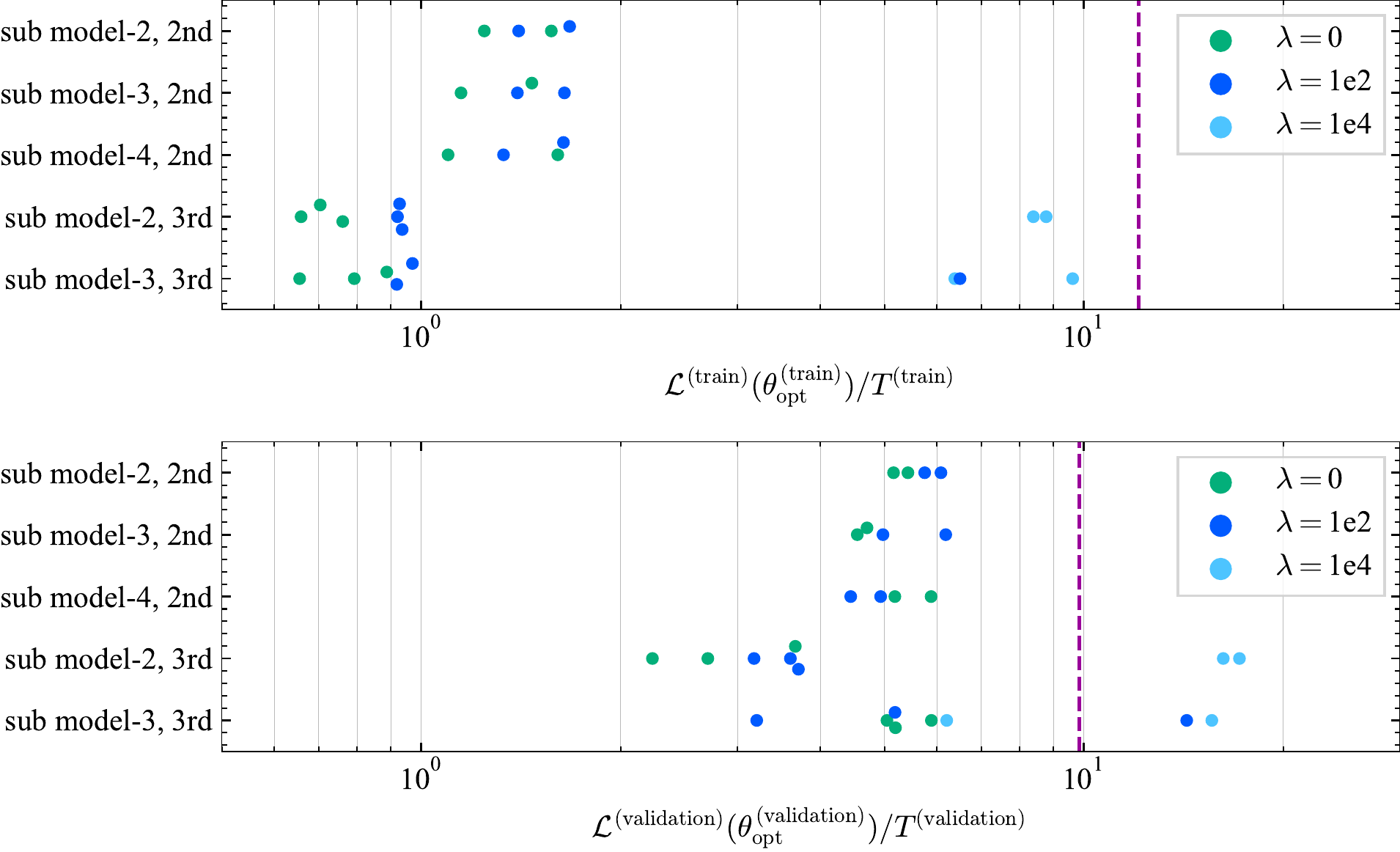}
            \subcaption{Color labeled by value of $\lambda$.}
            \label{fig:lambda_results}
    \end{minipage}
    \caption{ Relation between hyperparameters and estimation performance of the optimal model. The vertical purple dashed line represents MMG-EFD model results.}
    \label{fig:result_hyperparam}
\end{figure}

In the previous section, we discussed the effect of random seeding on performance estimation; we found that $\alpha$ and $\lambda$ show different trends depending on the dynamic model. Therefore, we next select the hyperparameters: $\alpha$ and $\lambda$, based on the performance on $\mathcal{D}\valid$. The error norm $\Ltrainprime$ and $\Lvalidprime$ of the optimal model for all computational cases defined in \Cref{tab:hypparam_2nd,tab:hypparam_3rd} are shown in \Cref{fig:result_hyperparam}. In \Cref{fig:alpha_results}, dot plots of $\Ltrainprime,~\Lvalidprime$ are color-coded by $\alpha$ values, and in \Cref{fig:lambda_results} dot plots are color-coded by $\lambda$ values.



For $\mathcal{D}\train$, the fitness was higher when the value of $\alpha$ and $\lambda$ were close to zero. For $\mathcal{D}\valid$, the same trend was seen for $\lambda$, but there was no clear trend due to the deviation penalty $\alpha$. Since the deviation penalty and regularization are intended to prevent overestimation of acceleration and overfitting, both caused by the complexity of the proposed model, the appropriate value is expected to depend on the characteristics of the dynamic model and dataset. However, we could not observe any clear performance degradation due to the penalties. Therefore, when using the proposed method, the user is recommended to perform parameter studies on the validation dataset for $\alpha$ and $\lambda$. For example, we introduced penalties to expect a regularization effect for models with high searching dimensions, but $\lambda=1e4$ has worse $\Ltrainprime,~\Lvalidprime$ are both worse than in the other cases, suggesting that the regularization penalty was too large.

In the next section, we selected the combination of $\alpha,~\lambda$ that minimize $\mathcal{L}\valid$ for each dynamic model, and compared the performance of each dynamic model with the test data.

\subsection{Evaluation by Test Data}\label{sec:result_test}
The generalization performance of the optimal models is checked on the test data.
As with training and validation, the following simplified notations are made for test and crash-astern: 
\begin{equation}\label{eq:notation_test_ca}
    \begin{aligned}
      \Ltestprime &=\mathcal{L}(\boldsymbol{\theta}\opt\valid; \mathcal{D}\test)/T\test \\
      \Lcaprime &=\mathcal{L}(\boldsymbol{\theta}\opt\valid; \mathcal{D}\ca)/T\ca \\
    \end{aligned}
\end{equation}

\subsubsection{Random Maneuvers}\label{sec:test_random}
\begin{figure}
    \centering
    \begin{minipage}[b]{\linewidth}
            \includegraphics[keepaspectratio, width=\columnwidth]{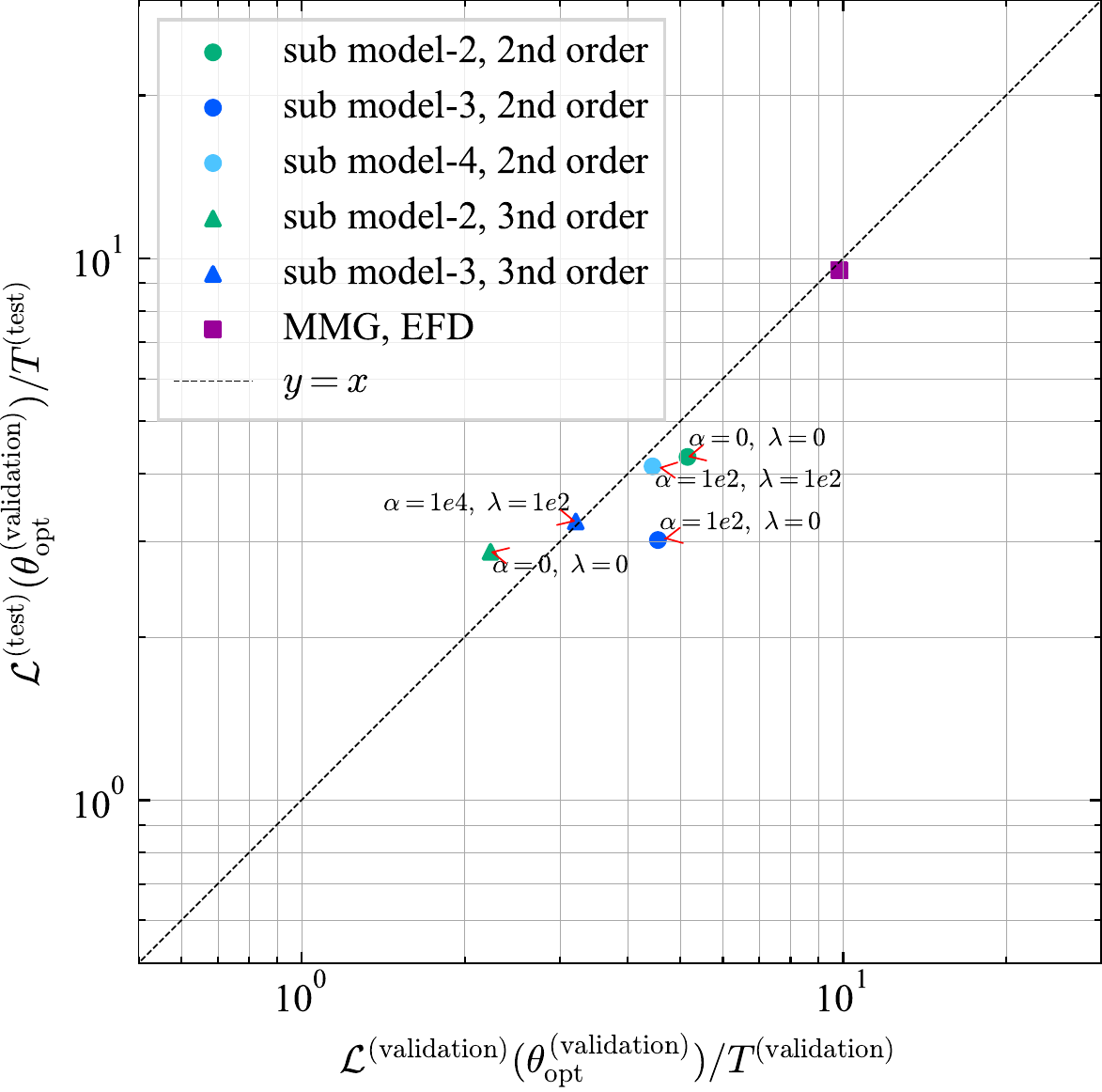}
            \subcaption{Random maneuver.}
            \label{fig:test}
    \end{minipage}
    \begin{minipage}[b]{\linewidth}
            \includegraphics[keepaspectratio, width=\columnwidth]{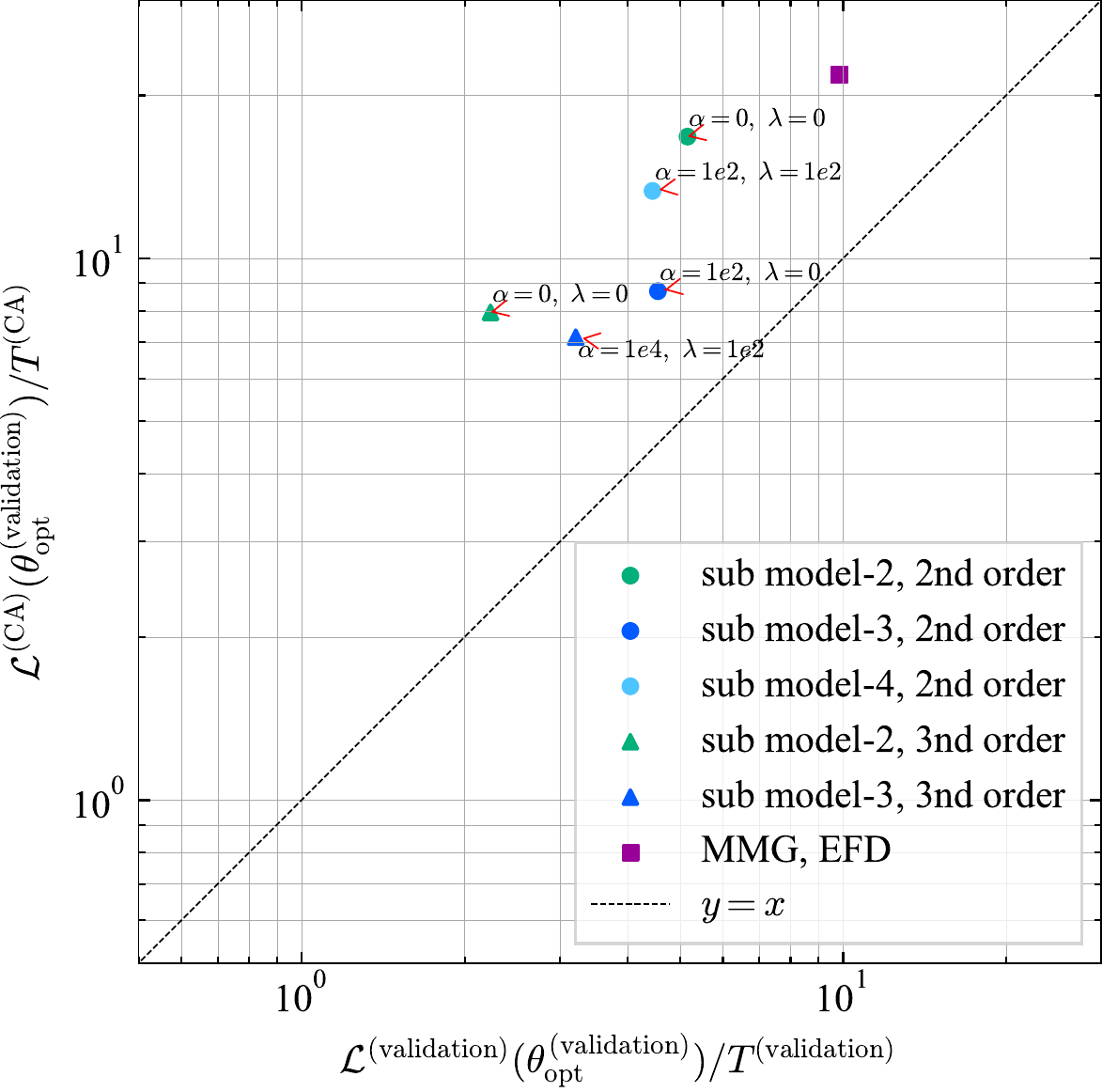}
            \subcaption{Crash-astern.}
            \label{fig:ca}
    \end{minipage}
    \caption{Performance on Test dataset $\mathcal{D}\test$ and $\mathcal{D}\ca$.}
    \label{fig:result_testdata}
\end{figure}

The Test data results for the optimal model for each dynamic model are shown in \Cref{fig:test}. The figure shows that all the optimal models outperform MMG-EFD in $\mathcal{D}\test$. The plots are generally on the linear line shown by the dashed line in the figure. In other words, the optimal parameter $\boldsymbol{\theta}\opt\valid$ chosen by the $\mathcal{D}\valid$ achieves the same accuracy with unknown test data. The MMG-EFD model is also on the same line, indicating that the proposed method has a generalization performance that can maintain its estimation accuracy regardless of the data set under random maneuvers, same as MMG-EFD model.

\begin{figure}
    \centering
    \begin{minipage}[t]{\linewidth}
            \centering         
            \includegraphics[keepaspectratio, width=\hsize, trim=70 70 40 0, clip]{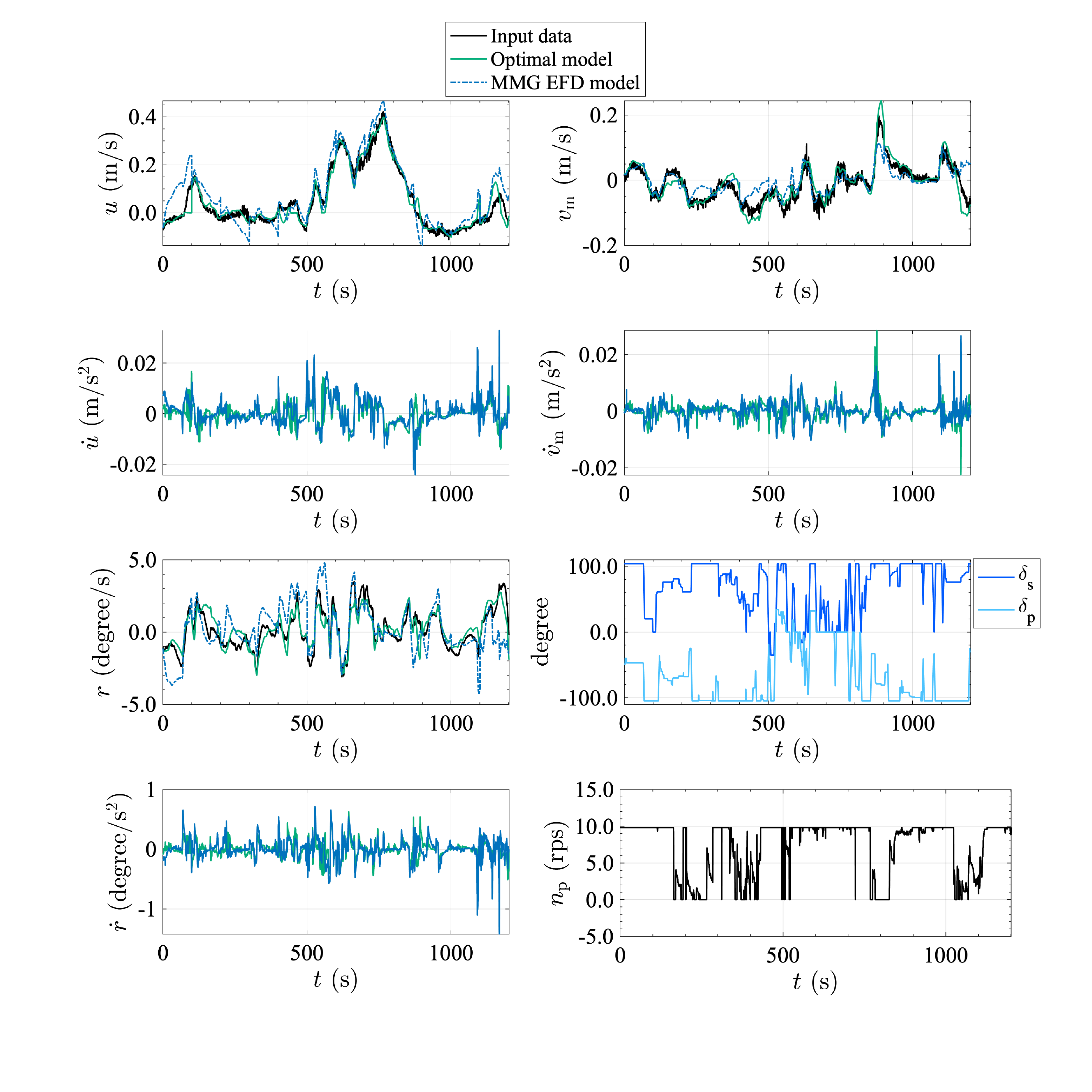}
            \subcaption{Best model; SM-2, 3rd order model, $\alpha= 0,~\lambda=0$.}
            \label{fig:best_test}
    \end{minipage}
    \begin{minipage}[t]{\linewidth}
    \centering
    \includegraphics[keepaspectratio, width=\hsize, trim=70 70 40 0, clip]{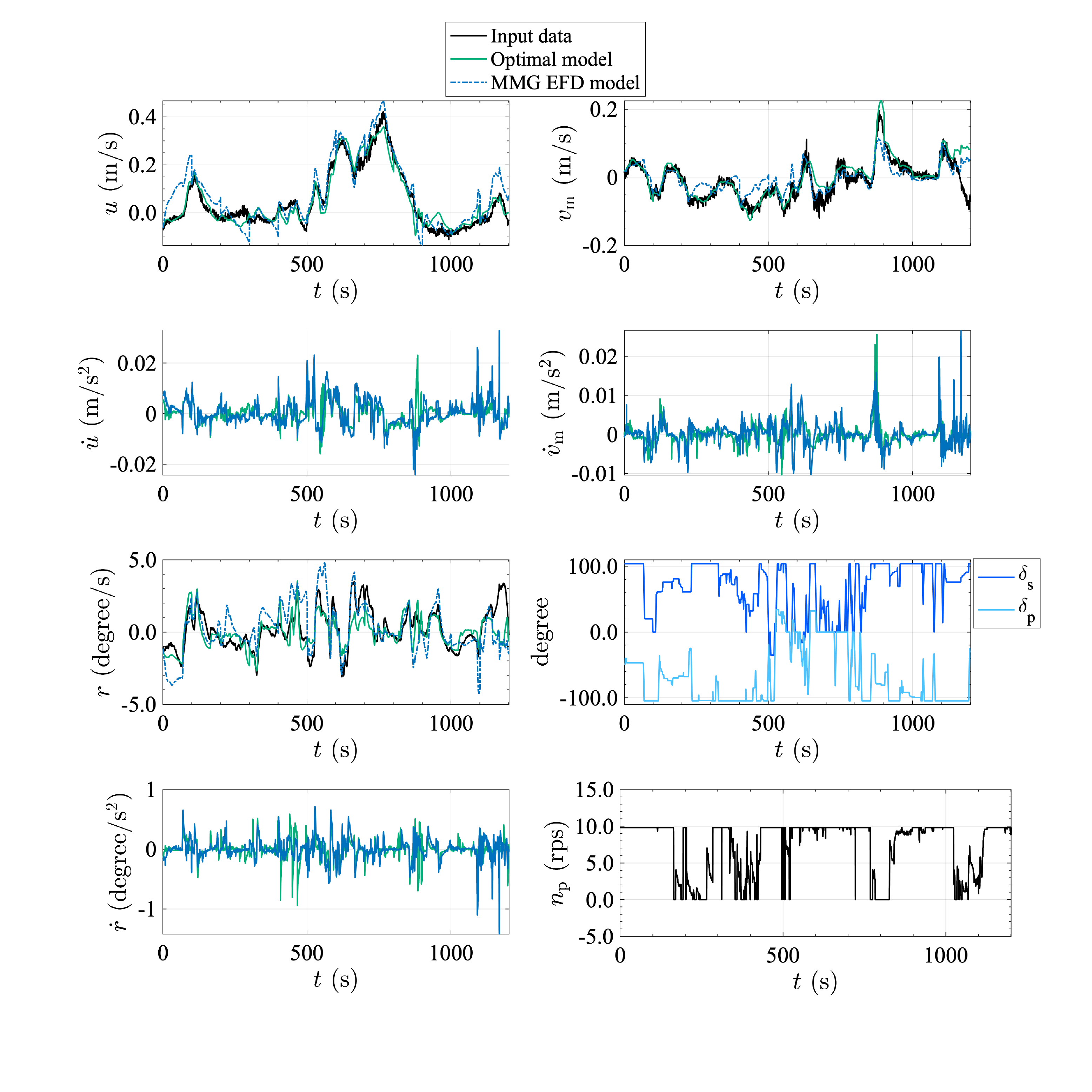}
            \subcaption{Worst model; SM-2, 2nd order model; $\alpha= 0,~\lambda = 0$.}
            \label{fig:worst_test}
    \end{minipage}
    \caption{Result of maneuvering simulation on $\mathcal{D}\test$ by optimal models of the best dynamic model and worst dynamic model. This figure shows time histories of $\boldsymbol{s}_{t}\simu$ and $\dot{\boldsymbol{s}}_{t}\simu$ of maneuvering simulation using optimal model and MMG-EFD model for $\mathcal{D}\test$. This figure also shows time histories of $\boldsymbol{a}_{t}$ and the measured state of the free-running model test $\boldsymbol{s}_{t}\test$.}
    \label{fig:test_timehist}
\end{figure}

Next, we show the results of the maneuvering simulation using the best optimal model and the worst optimal model to verify the performance of the proposed model in detail. Based on the \Cref{fig:test}, SM-2, 2nd order model, $\alpha=0,~\lambda=0$, is the worst, and the SM-2, 3rd order model, $\alpha=0,~\lambda=0$ is the best.
The maneuvering simulation results of the two models are shown in \cref{fig:test_timehist}. The simulation results of the MMG-EFD model and the time series of $\mathcal{D}\test$ are also compared. The simulation results from both optimal models agreed with the experimental results. In addition, the acceleration $\boldsymbol{\dot{s}}_{t}=(\dot{u},~\dot{v}_{\mathrm{m}},~\dot{r})$ of the optimal model and the MMG-EFD model are of the same order. In other words, the proposed model with optimal parameters can obtain the same level of acceleration as the MMG-EFD model, which is a model constructed from hydrodynamic forces measured by model tests, even though the acceleration is not included in the optimization of the proposed model.

\subsubsection{Crash-Astern}\label{sec:test_ca}
Here we validate the proposed model for maneuvers other than random maneuvers using the crash-astern (CA) test data set $\mathcal{D}\ca$. The error norm for each dynamic model is shown in \Cref{fig:ca}. The correlation coefficient between $\Lvalidprime$ and $\Lcaprime$ is $r=0.76$ and positively correlated. In other words, by improving the performance on random maneuvers, performance could be improved even for data with different types of maneuvers. As shown in \cref{fig:hist_train_vs_ca}, the distributions of $\mathcal{D}\train$ and $\mathcal{D}\ca$ are different. Consequently, the proposed method can improve the estimation performance for various motions and generate a stable dynamic model. Compared to $\mathcal{D}\test$, $\mathcal{L}$ increased by a factor of 2 to 3, which means the estimation accuracy worsened, but the MMG-EFD model also showed a similar degree of deterioration.



\begin{figure}
    \centering
    \begin{minipage}[b]{\linewidth}
            \includegraphics[keepaspectratio, width=\hsize, trim=20 0 0 80, clip]{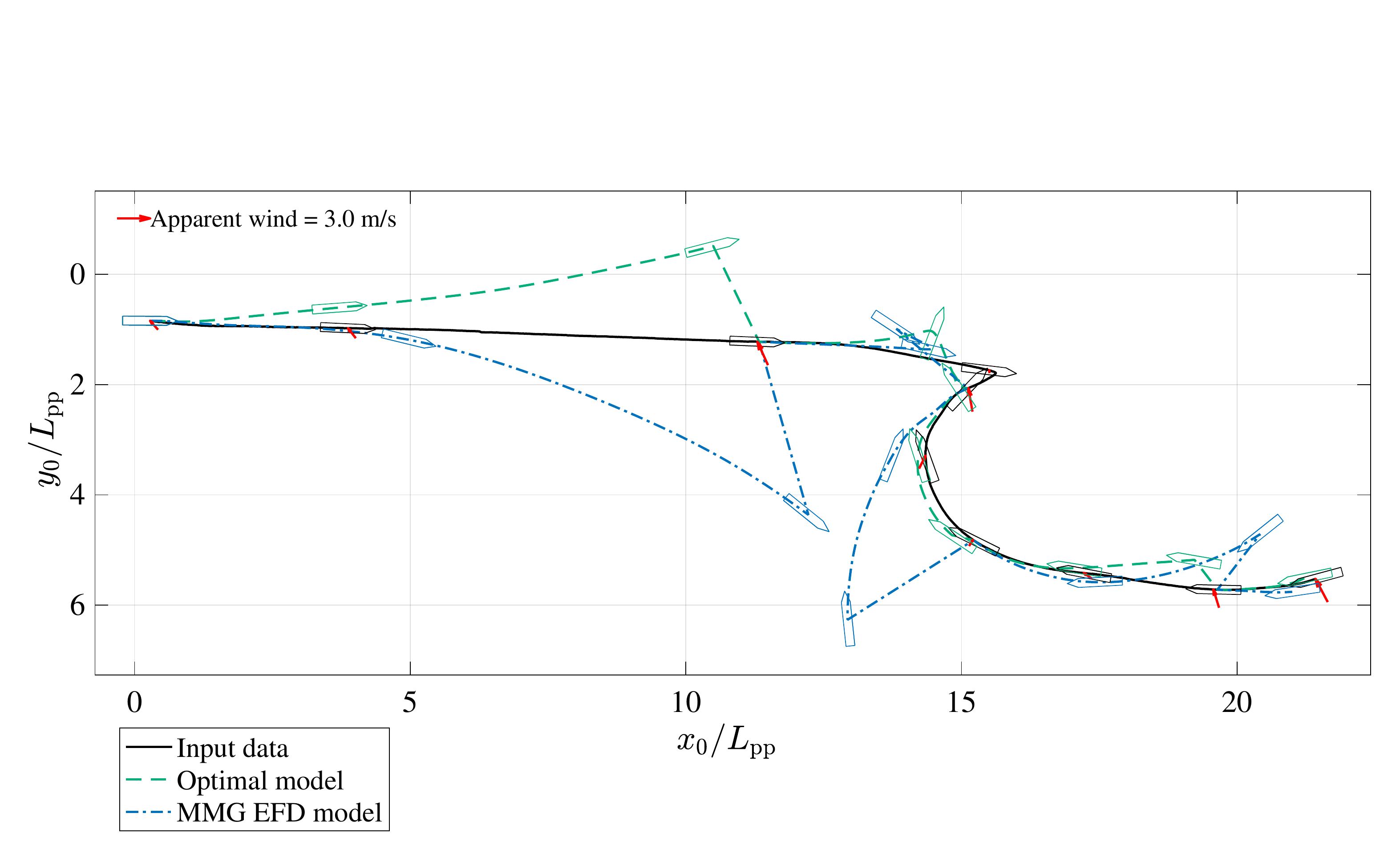}
            \subcaption{Best model; SM-2, 3rd order model; $\alpha= 0,~\lambda = 0$.}
            \label{fig:best_traj}
    \end{minipage}
    \begin{minipage}[b]{\linewidth}
            \includegraphics[keepaspectratio, width=\hsize, trim=20 0 0 80, clip]{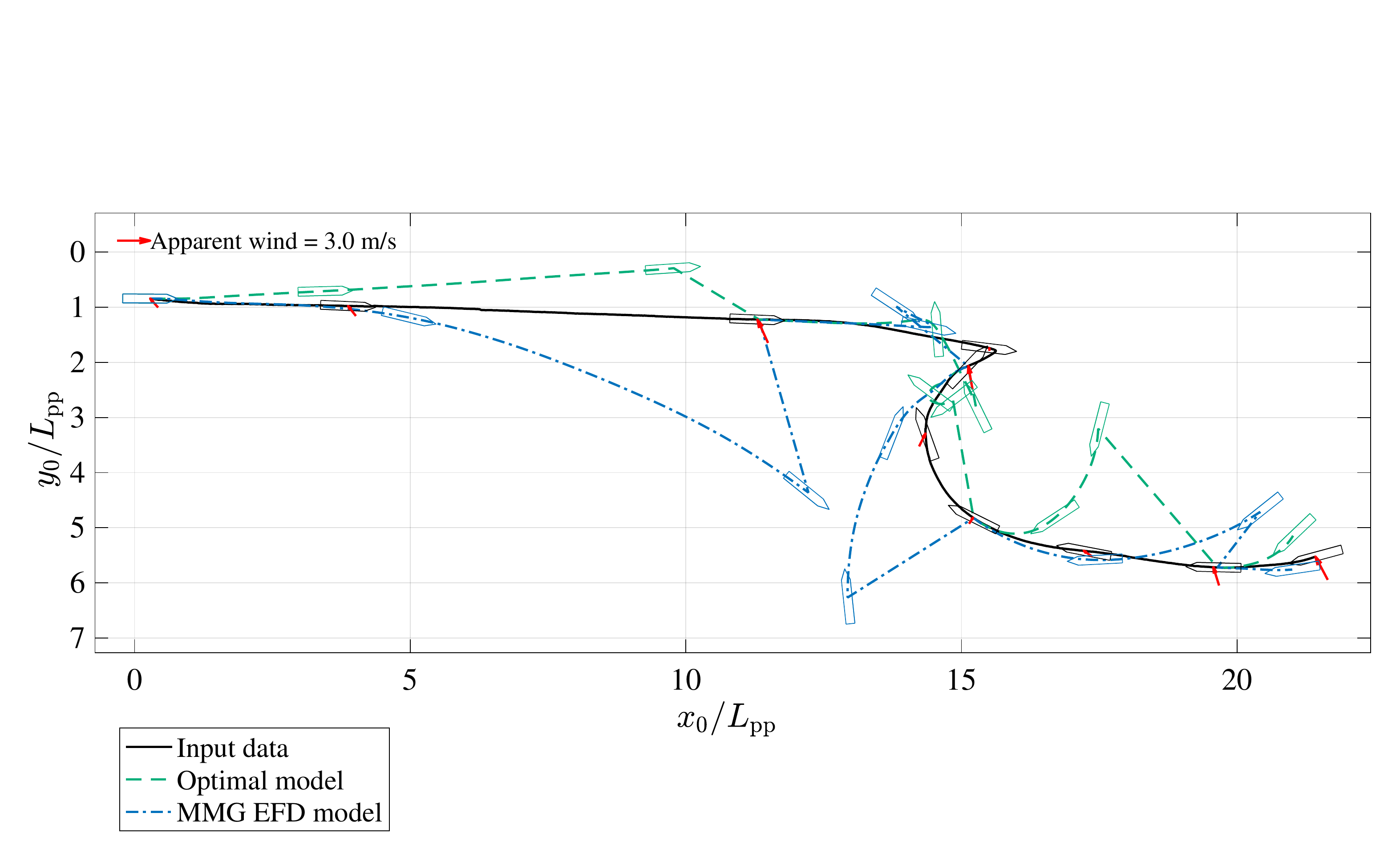}
            \subcaption{Worst model; SM-2, 2nd order model; $\alpha= 0,~\lambda = 0$.}
            \label{fig:worst_traj}
    \end{minipage}
    \caption{Estimated trajectories of maneuvering simulation using two optimal models and MMG-EFD model for $\mathcal{D}\ca$. Ship-like pentagons represent ship positions and headings at CS's beginning, middle, and end. The black line represents the input dataset, i.e., measured trajectory of the free-running model test. Optimal model is SM-2, 2nd order, $\alpha=0,~\lambda=0$. }
    \label{fig:ca_traj}
\end{figure}

\begin{figure}
    \centering
    \begin{minipage}[b]{\linewidth}
            \centering
            \includegraphics[keepaspectratio, width=\hsize, trim=70 70 40 0, clip]{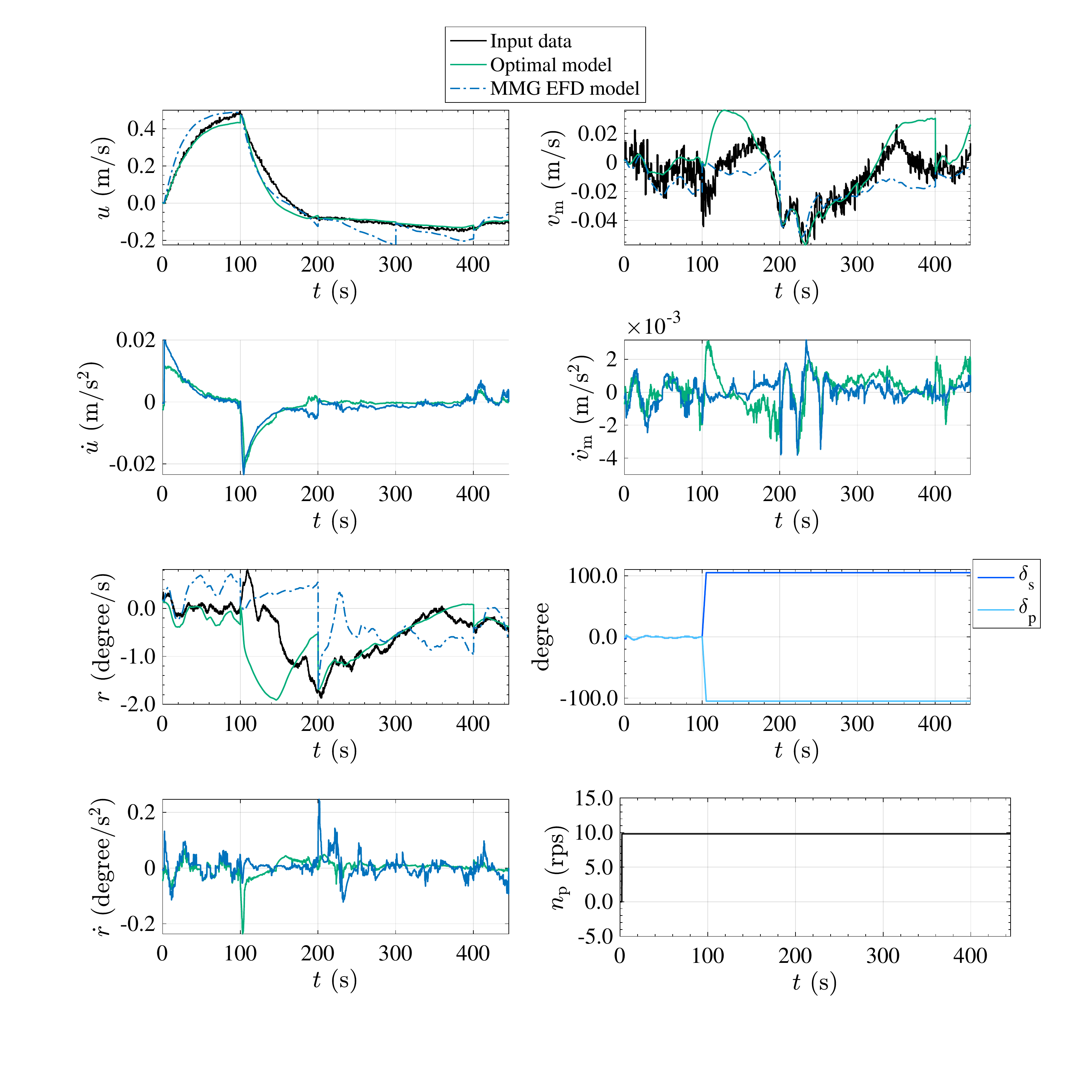}
            \subcaption{Best model; SM-2, 3rd order model; $\alpha= 0,~\lambda = 0$.}
            \label{fig:best_hist}
    \end{minipage}\\
    \begin{minipage}[b]{\linewidth}
            \centering
            \includegraphics[keepaspectratio, width=\hsize, trim=70 70 40 0, clip]{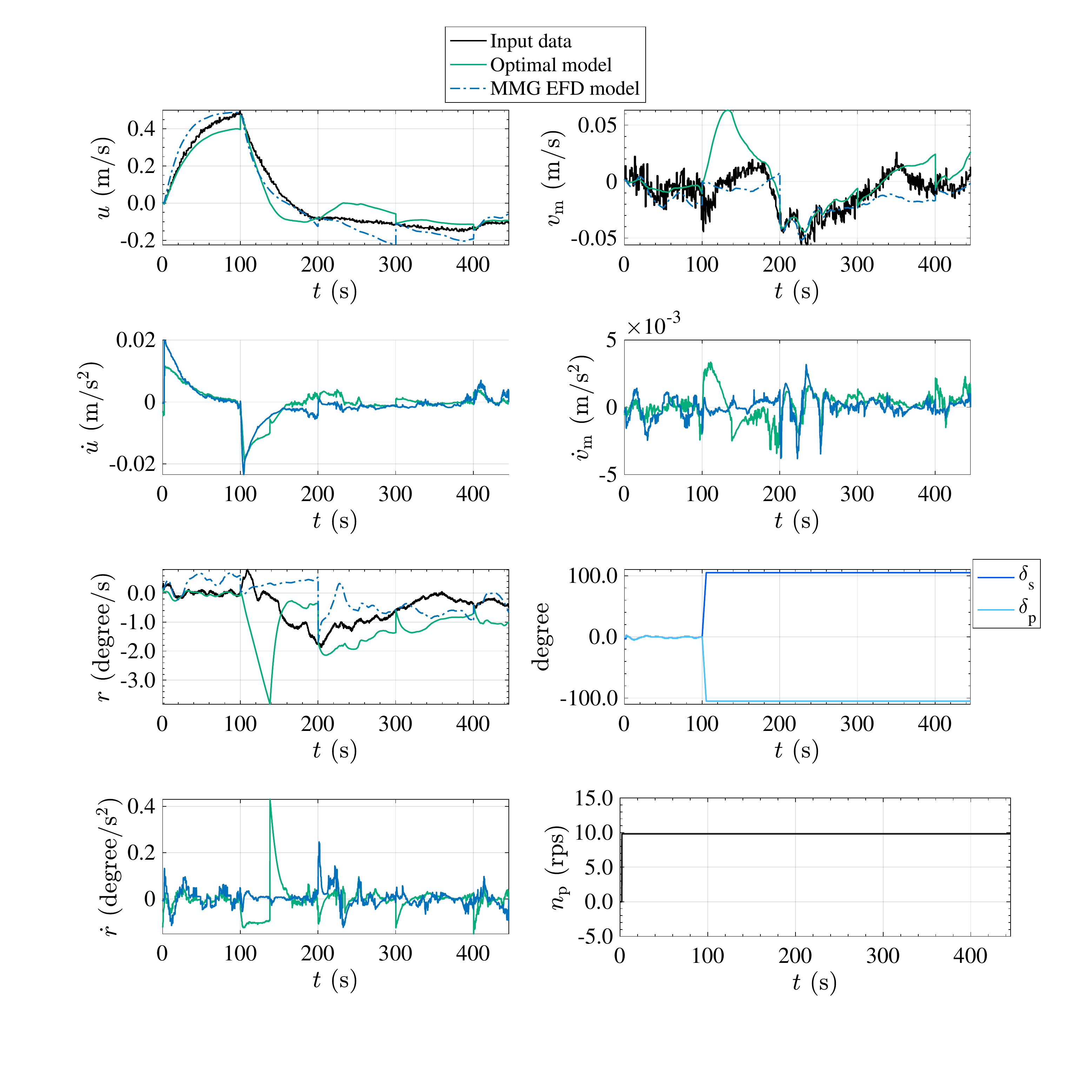}
            \subcaption{Worst model; SM-2, 2nd order model; $\alpha= 0,~\lambda = 0$.}
            \label{fig:worst_hist}
    \end{minipage}
    \caption{Time histories of $\boldsymbol{s}_{t}\simu$ and $\dot{\boldsymbol{s}}_{t}\simu$ of maneuvering simulation using best optimal model and MMG-EFD model for $\mathcal{D}\ca$. This figure also shows time histories of control inputs and $\boldsymbol{s}_{t}\ca$, i.e., measured state of free-running model test. Optimal model is SM-2, 2nd order, $\alpha=0,~\lambda=0$.}
    \label{fig:ca_hist}
\end{figure}

The time series of the maneuvering simulation with the best model are compared with the physical experiment results. The results of maneuvering simulation using the best and worst model for $\mathcal{D}\ca$ are shown in \Cref{fig:ca_traj,fig:ca_hist}. In \Cref{fig:ca_traj,fig:ca_hist}, one of the four CA tests included in $\mathcal{D}\ca$ is shown. This particular CA test has the smallest $\Lcaprime$ with simulation using the best model.
The trajectory $(x_{0},~y_{0},~\psi)$ shown in \Cref{fig:ca_traj} agrees with the experimental trajectory well except for $\psi$ immediately after the start of the stopping maneuver in both optimal models and during astern in the worst model. In addition, the time series of $\boldsymbol{s}_{t}\simu$ and $\dot{\boldsymbol{s}}_{t}\simu$shown in \Cref{fig:ca_hist}. In both of best and worst model's results, the estimation performance of time-series showed two types of trend on contiguous subsequences (CS); CS with a good agreement and CS with a large discrepancy. In particular, the deviations were considerable for the CS at $100<t<200$ s immediately after the start of the crash-astern steering. Moreover, slopes of $\vm,~r$ had opposite signs with the experiment at $100<t<120$ s. The velocity slope, i.e., acceleration, is not resolved correctly, meaning transient characteristics of hydrodynamic forces are not represented at that period. In this period, $u=0.5$ m/s at $t=0$ of the subsequence, which is the velocity extrapolated from the training data as shown in \Cref{fig:hist_train_vs_ca}. The extrapolation of the training data is considered to have degraded the estimation performance.

\subsubsection{Analysis on Error}
 This section analyzes the distribution of errors in the maneuvering simulations for each data set, to summarize the estimation performance on the training, validation, and test-CA datasets. The model used is the best model, SM-2, 3rd order model, $\alpha=\lambda=0$. Since the error function \Cref{eq:norm_def,eq:valid_norm_def} is the sum of errors, the squared error vector $\boldsymbol{\varepsilon}_{t}\equiv \big(\varepsilon_{t,u},~\varepsilon_{t,\vm},~\varepsilon_{t,r}\big)\in\R^{3}$ at time $t$ is analyzed. Here, the following equation defines the $j$-th component of $\boldsymbol{\varepsilon}_{t}$ as follows,
\begin{equation}
    \varepsilon_{t,j} = \big\{\hat{s}_{t,j}^{(\text{input},i)}- \hat{s}_{t,j}^{(\text{sim},i)}(\boldsymbol{\theta}_{\text{opt}};~\mathcal{D}^{(\text{input})})\big\}^{2}\enspace.
\end{equation}
Here, $\hat{s}_{t,j}^{(\cdot,i)}$ is the $j$-th component of the standardized state variables at time $t$ of the $i$-th CS. The standardized state variables are $\hat{\boldsymbol{s}}_{t}^{(\cdot,i)}=\big(\hat{u}^{(\cdot,i)}(t),~\hat{v}_{\mathrm{m}}^{(\cdot,i)}(t),~\hat{r}^{(\cdot,i)}(t))$.

\begin{figure}
    \centering
    \begin{minipage}[t]{\linewidth}
            \includegraphics[keepaspectratio, width=\columnwidth]{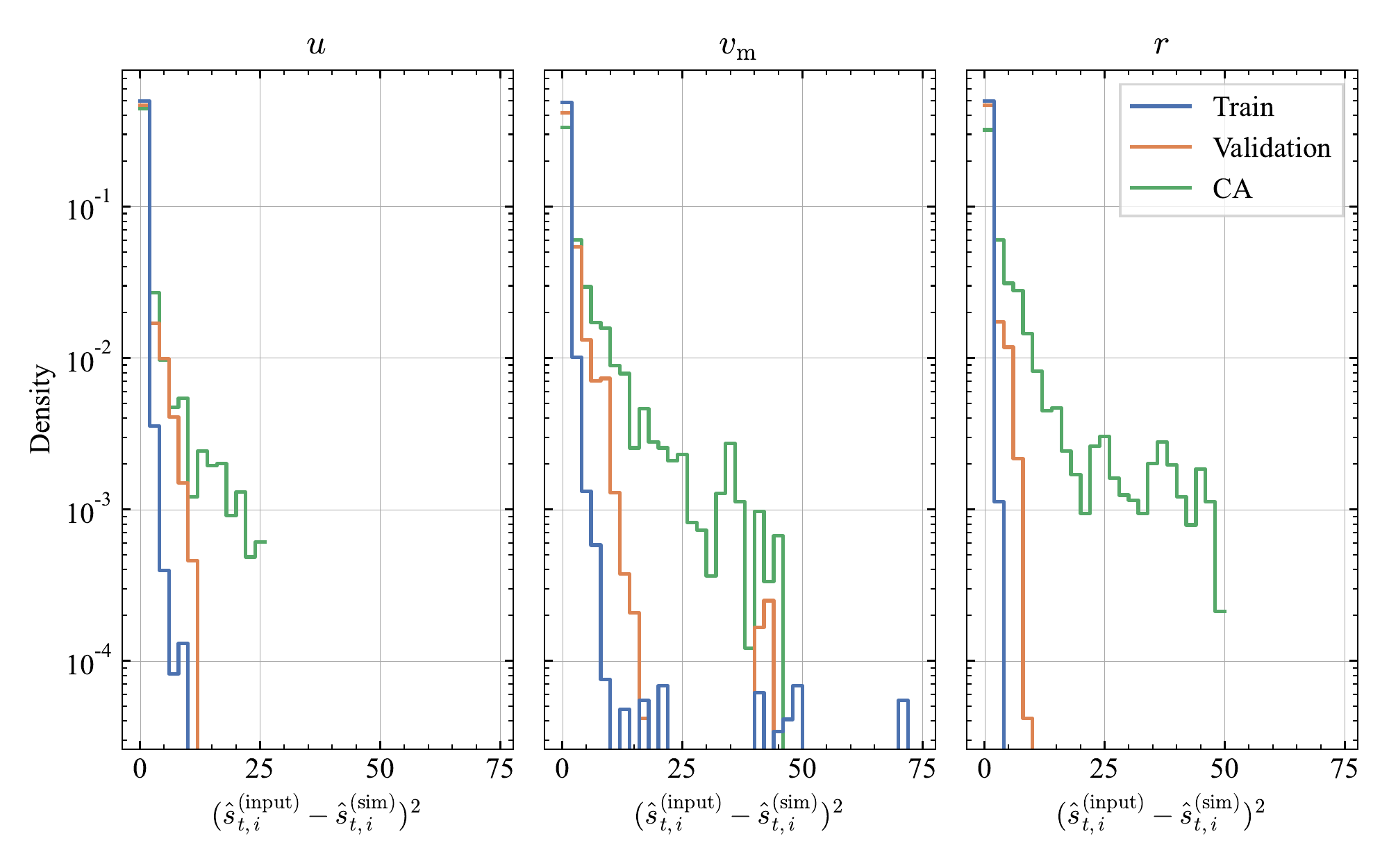}
            \subcaption{Histogram of instantaneous errors in log scale vertical axis.}
            \label{fig:error_ylog}
    \end{minipage}\\
    \begin{minipage}[t]{\linewidth}
            \includegraphics[keepaspectratio, width=\columnwidth]{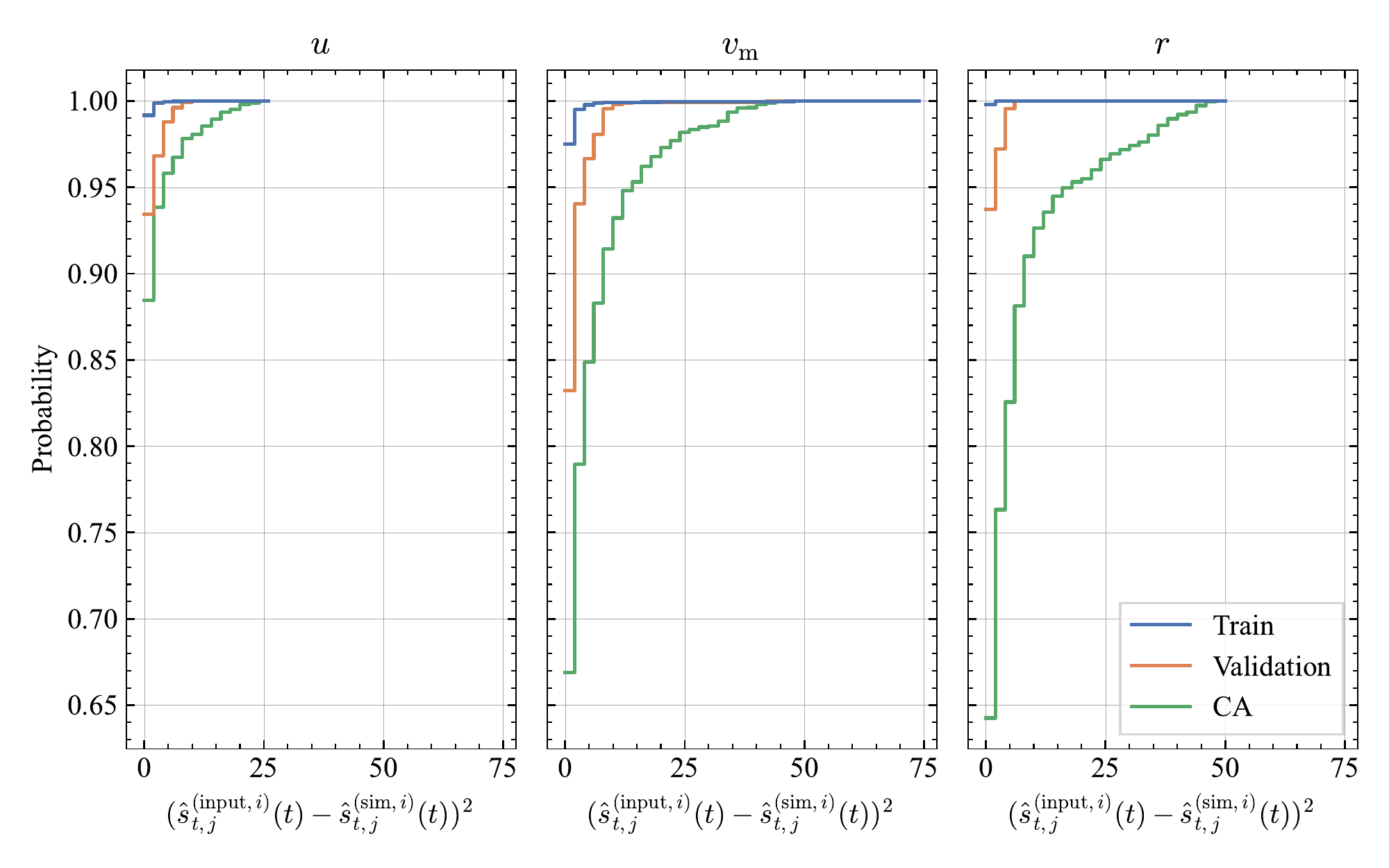}
            \subcaption{Cumulative histogram of errors.}
            \label{fig:error_cumsum}
    \end{minipage}
    \caption{Histograms of instantaneous error $\boldsymbol{\varepsilon}_{t} =  \big(\varepsilon_{t,u}, ~\varepsilon_{t,\vm},~\varepsilon_{t,r}\big)$ per data sets: $\mathcal{D}\train,~\mathcal{D}\valid$ and $\mathcal{D}\ca$. }
    \label{fig:errors}
\end{figure}

The histogram of $\boldsymbol{\varepsilon}_{t}$ is shown in \Cref{fig:error_ylog} and the cumulative frequency histogram of $\boldsymbol{\varepsilon}_{t}$ is shown in \Cref{fig:error_cumsum}. The figure shows that test-CA contains larger instantaneous errors than training and validation. In test-CA, $\vm$ and $r$ have longer legs of the distribution than $u$, which means $\vm$ and $r$ contain larger instantaneous errors. In addition, cumulative frequencies of $\varepsilon_{t,\vm},~\varepsilon_{t,r}$ increased gradually for $\varepsilon_{t,j}>15$. From this result, we can say that a specific large instantaneous error does not increase the sum of errors; instantaneous errors of about 5 to 10$\%$ of the overall time steps are worsening. Therefore, measures to improve the accuracy of these lower 5 to 10$\%$ time steps of $\vm$ and $r$ are needed. 

\section{Discussion}\label{sec:discussion}




\paragraph{\AddD{Modeling Automation}}
\AddD{Before concluding the study, we explain how the proposed model and method enable modeling automation. First, because of the simple rules for model derivation, derivation can be automated. In contrast, existing hydrodynamic models must derive the expression based on hydrodynamics for each actuator configuration. Second, the necessary inputs for the proposed method are only the dataset. Generation of datasets itself can not be automated with the proposed method; however, a priori hydrodynamic information other than dataset is not necessary. Since CMA-ES with the restart strategy is a quasi-parameter-free optimization \citep{Auger2005}, the user is not required to seek the parameter for the optimization itself. Third, selection of the model formulae can also be automated because multiple model formulae can be derived automatically and selected by the evaluation based on maneuvering simulation. Because of those three reasons, the proposed method can essentially remove the human work required to generate dynamic models. By employing the proposed method, we can develop an algorithm that automatically generates a dynamic model suitable for harbor maneuvers once the user feeds a dataset.}




\paragraph{Limitations and Future work}

The main drawback of the proposed method is the computation time required for optimization. The SM-2, 2nd order model, a model with the smallest number of parameters, requires three days of computation time, and the SM-3, 3rd order model, the largest number of parameters, requires ten days. The computation conditions were 16 parallel Intel Xeon Platinum 8260 computers, and the language and compiler were Fortran 90 and Intel Fortran. CMA-ES is easy to parallelize and may be faster depending on machine power, but it is computationally expensive in general because of the need for parallel-capable machines.

A remaining issue is a dependence on the dataset size. In this study, the dataset size was kept constant. A total of approximately 10,000 seconds of random maneuvers was used. If $\lpp=100$m, this corresponds to about 16 hours. As the size of the target vessel increases, the cost of data acquisition also increases. Therefore, in addition to the dependence on the amount of data, reducing the amount of data required and efficient methods of data acquisition are also future work. In addition, introducing non-dimensionalization without velocity $U$, such as \citep{Kose1985}, may ease the domain boundary setting of coefficients.

\section{Conclusion}\label{sec:conc}
In this study, we proposed the \EraseD{novel }\AddD{Abkowitz-MMG hybrid}\EraseD{mathematical maneuvering} model and its parameter identification method \AddD{for harbor maneuvers and }to realize automated modeling\EraseD{ of berthing maneuvers}.

The proposed model can be derived according to \EraseD{monolithic}\AddD{simple} rules using Taylor expansion and has a high degree of freedom to express \AddD{harbor maneuvers'} complex motions and ship handlings\EraseD{ berthing and unberthing and ship handling in port}. The proposed model can be \AddD{more easily} derived \AddD{than existing Hydrodynamic models for arbitrary ship configurations.} \EraseD{according to simple rules, even if the ship configuration or type of motion changes.} \AddD{In addition, the physical meaning of the model's formulae and their terms is much more understandable than those using neural networks.}

\AddD{We also proposed a method to identify the model parameters and select the model's formulae using System Identification. Even though the proposed model needs to identify several hundred model parameters because of its simple rules for derivation,  SI using CMA-ES enables identifying model parameters within reasonable computational time. Moreover, }
since multiple models can be easily derived, we identified the parameters of several models and selected the best model from them. Thus, the proposed method does not depend on the captive model tests and knowledge of ship hydrodynamics to select appropriate mathematical expressions of the model, thereby reducing the amount of labor required for model selection. In addition, simple rule-based model derivation and easy model selection methods can relax the necessary skill and facility requirements for users to perform model generation.

This study used trajectories of a free-running model ship of a single-propeller, VecTwin-rudder-equipped ship, as the data set. As a result, we confirmed that even a dynamic model derived from simple rules could estimate unknown harbor maneuvers equivalently or more accurately than a dynamic model generated by existing methods.

\AddD{In summary, we showed that the proposed model and method could reduce human intervention in model generation and have adequate estimation performance. Consequently, they can enable automated modeling, which automates the modeling process itself. Although design and development of the automation algorithm are the remaining tasks, the basic concept of modeling automation, and the core of the automation algorithm, i.e., the dynamic model and the model generation method, were proposed in this study.}

\section*{Acknowledgements}
This study was conducted as a part of the Nippon Foundation's ``demonstration tests of The Nippon
Foundation MEGURI2040 Fully Autonomous Ship Program.'' This study was also supported by a Grant-in-Aid for Scientific Research from the Japan Society for Promotion of Science (JSPS KAKENHI Grant \#22H01701). This study was conducted under a joint research project between Japan Hamworthy Co., Ltd. and Osaka university.


The authors would like to express gratitude to Prof. Masaaki Sano at Hiroshima University, for providing captive model test results. The authors also would like to thank students of the Ship Intelligentization Subarea, Osaka University, for supporting the model experiment: Dimas M. Rachman, Nozomi Amano, Hiroaki Koike, and Yuta Fueki.
\bibliographystyle{spphys}
\bibliography{main_en}

\end{document}

%% file: xderivative.tex
\boldsymbol{X}^{(i=1\cdots m)} =
\big( X^{(i)}_{u},~X^{(i)}_{\cos(\deltas)},~X^{(i)}_{\cos(\deltap)},~X^{(i)}_{\np},~
        X^{(i)}_{u|u|},~X^{(i)}_{u\cos(\deltas)},~
        X^{(i)}_{u\cos(\deltap)},~
        X^{(i)}_{u\np},~
        X^{(i)}_{v\subm v\subm },~
        X^{(i)}_{v\subm r},~
        X^{(i)}_{v\subm \sin(\deltas)},~
        X^{(i)}_{v\subm \sin(\deltap)},~
        X^{(i)}_{rr},~
        X^{(i)}_{r\sin(\deltas)},~
        X^{(i)}_{r\sin(\deltap)},~
        X^{(i)}_{\sin(\deltas)\sin(\deltap)},~
        X^{(i)}_{\cos(\deltas)\cos(\deltap)},~
        X^{(i)}_{\cos(\deltas)\np},~
        X^{(i)}_{\cos(\deltap)\np},~X^{(i)}_{\np|\np|}, 
        X^{(i)}_{uuu},~X^{(i)}_{uv\subm v\subm },~X^{(i)}_{uv\subm r},~
        X^{(i)}_{urr},~
        X^{(i)}_{uv\subm \sin(\deltas)},~
        X^{(i)}_{ur\sin(\deltas)},~
        X^{(i)}_{u|u|\cos(\deltas)},~
        X^{(i)}_{v\subm v\subm \cos(\deltas)},~
        X^{(i)}_{v\subm r\cos(\deltas)},~
        X^{(i)}_{v\subm \sin(\deltas)\cos(\deltas)},~
        X^{(i)}_{rr\cos(\deltas)},~
        X^{(i)}_{uv\subm \sin(\deltap)},~
        X^{(i)}_{ur\sin(\deltap)},~
        X^{(i)}_{u\sin(\deltas)\sin(\deltap)},~
        X^{(i)}_{v\subm \cos(\deltas)\sin(\deltap)},~
        X^{(i)}_{r\cos(\deltas)\sin(\deltap)},~
        X^{(i)}_{r\sin(\deltas)\cos(\deltas)},~
        X^{(i)}_{\sin(\deltas)\cos(\deltas)\sin(\deltap)},~
        X^{(i)}_{u|u|\cos(\deltap)},~
        X^{(i)}_{u\cos(\deltas)\cos(\deltap)},~
        X^{(i)}_{v\subm v\subm \cos(\deltap)},~
        X^{(i)}_{v\subm r\cos(\deltap)},~
        X^{(i)}_{v\subm \sin(\deltas)\cos(\deltap)},~
        X^{(i)}_{v\subm \sin(\deltap)\cos(\deltap)},~
        X^{(i)}_{rr\cos(\deltap)},~
        X^{(i)}_{r\sin(\deltas)\cos(\deltap)},~
        X^{(i)}_{r\sin(\deltap)\cos(\deltap)},~
        X^{(i)}_{\sin(\deltas)\sin(\deltap)\cos(\deltap)},~
        X^{(i)}_{u|u|\np},~X^{(i)}_{u\cos(\deltas)\np},~
        X^{(i)}_{u\cos(\deltap)\np},~
        X^{(i)}_{u\np|\np|},~
        X^{(i)}_{v\subm v\subm \np},~
        X^{(i)}_{v\subm r\np},
        X^{(i)}_{v\subm \sin(\deltas)\np},~
        X^{(i)}_{v\subm \sin(\deltap)\np},~
        X^{(i)}_{rr\np},~
        X^{(i)}_{r\sin(\deltas)\np},~
        X^{(i)}_{r\sin(\deltap)\np},
        X^{(i)}_{\sin(\deltas)\sin(\deltap)\np},~
        X^{(i)}_{\cos(\deltas)\cos(\deltap)\np},~
        X^{(i)}_{\cos(\deltas)\np|\np|},~
        X^{(i)}_{\cos(\deltap)\np|\np|},~
        X^{(i)}_{\np\np\np} \big)

%% file: z_even.tex
\boldsymbol{z}_{\text{even}} =  
\big( u,~\cos(\deltas),~\cos(\deltap),~\np,~ 
    u|u|,~
    u\cos(\deltas),~
    u\cos(\deltap),~
    u\np,~
    v\subm v\subm ,~
    v\subm r,~
    v\subm \sin(\deltas),~
    v\subm \sin(\deltap),~
    rr,~
    r\sin(\deltas),~
    r\sin(\deltap),~
    \sin(\deltas)\sin(\deltap),~
    \cos(\deltas)\cos(\deltap),~
    \cos(\deltas)\np,~
    \cos(\deltap)\np,~
    \np|\np|,~
    uuu,~uv\subm v\subm ,~uv\subm r,~urr,~uv\subm \sin(\deltas),~
    ur\sin(\deltas),~
    u|u|\cos(\deltas),~
    v\subm v\subm \cos(\deltas),~
    v\subm r\cos(\deltas),~
    v\subm \sin(\deltas)\cos(\deltas),~
    rr\cos(\deltas),~
    r\sin(\deltas)\cos(\deltas),~
    uv\subm \sin(\deltap),~
    ur\sin(\deltap),~
    u\sin(\deltas)\sin(\deltap),~
    v\subm \cos(\deltas)\sin(\deltap),~
    r\cos(\deltas)\sin(\deltap),~
    \sin(\deltas)\cos(\deltas)\sin(\deltap),~
    u|u|\cos(\deltap),~
    u\cos(\deltas)\cos(\deltap),~
    v\subm v\subm \cos(\deltap),~
    v\subm r\cos(\deltap),~
    v\subm \sin(\deltas)\cos(\deltap),~
    v\subm \sin(\deltap)\cos(\deltap),~
    rr\cos(\deltap),~
    r\sin(\deltas)\cos(\deltap),~
    r\sin(\deltap)\cos(\deltap),~
    \sin(\deltas)\sin(\deltap)\cos(\deltap),~
    u|u|\np,~
    u\cos(\deltas)\np,~
    u\cos(\deltap)\np,~
    u\np|\np|,~
    v\subm v\subm \np,~
    v\subm r\np,~
    v\subm \sin(\deltas)\np,~
    v\subm \sin(\deltap)\np,~
    rr\np,~
    r\sin(\deltas)\np,~
    r\sin(\deltap)\np,~
    \sin(\deltas)\sin(\deltap)\np,~
    \cos(\deltas)\cos(\deltap)\np,~
    \cos(\deltas)\np|\np|,~
    \cos(\deltap)\np|\np|,~
    \np\np\np \big)

%% file: yderivative.tex
    \boldsymbol{Y}^{(i=1\cdots m)} = 
    \big( Y^{(i)}_{v\subm },~Y^{(i)}_{r},~Y^{(i)}_{\sin(\deltas)},~
    Y^{(i)}_{\sin(\deltap)},~
    Y^{(i)}_{uv\subm },~
    Y^{(i)}_{ur},~
    Y^{(i)}_{u\sin(\deltas)},~
    Y^{(i)}_{u\sin(\deltap)},~
    Y^{(i)}_{v\subm \cos(\deltas)},~
    Y^{(i)}_{v\subm \cos(\deltap)},~
    Y^{(i)}_{v\subm \np},~
    Y^{(i)}_{r\cos(\deltas)},~
    Y^{(i)}_{r\cos(\deltap)},~
    Y^{(i)}_{r\np},~
    Y^{(i)}_{\sin(\deltas)\cos(\deltas)},~
    Y^{(i)}_{\sin(\deltas)\cos(\deltap)},~
    Y^{(i)}_{\sin(\deltas)\np},~
    Y^{(i)}_{\cos(\deltas)\sin(\deltap)},~
    Y^{(i)}_{\sin(\deltap)\cos(\deltap)},~
    Y^{(i)}_{\sin(\deltap)\np},~
    Y^{(i)}_{uuv\subm },~Y^{(i)}_{v\subm v\subm v\subm },~Y^{(i)}_{uur},~Y^{(i)}_{v\subm |v\subm |r},~Y^{(i)}_{v\subm r|r|},~
    Y^{(i)}_{rrr},~
    Y^{(i)}_{uu\sin(\deltas)},~
    Y^{(i)}_{v\subm |v\subm |\sin(\deltas)},~
    Y^{(i)}_{v\subm r\sin(\deltas)},~
    Y^{(i)}_{r|r|\sin(\deltas)},~
    Y^{(i)}_{uv\subm \cos(\deltas)},~
    Y^{(i)}_{ur\cos(\deltas)},~
    Y^{(i)}_{u\sin(\deltas)\cos(\deltas)},~
    Y^{(i)}_{v\subm \cos(\deltas)\cos(\deltas)},~
    Y^{(i)}_{r\cos(\deltas)\cos(\deltas)},
    Y^{(i)}_{\sin(\deltas)\cos(\deltas)\cos(\deltas)},~
    Y^{(i)}_{uu\sin(\deltap)},~
    Y^{(i)}_{u\cos(\deltas)\sin(\deltap)},~
    Y^{(i)}_{v\subm |v\subm |\sin(\deltap)},~
    Y^{(i)}_{v\subm r\sin(\deltap)},~
    Y^{(i)}_{v\subm \sin(\deltas)\sin(\deltap)},~
    Y^{(i)}_{r|r|\sin(\deltap)},~
    Y^{(i)}_{r\sin(\deltas)\sin(\deltap)},~
    Y^{(i)}_{\cos(\deltas)\cos(\deltas)\sin(\deltap)},~
    Y^{(i)}_{uv\subm \cos(\deltap)},~
    Y^{(i)}_{ur\cos(\deltap)},~
    Y^{(i)}_{u\sin(\deltas)\cos(\deltap)},~
    Y^{(i)}_{u\sin(\deltap)\cos(\deltap)},~
    Y^{(i)}_{v\subm \cos(\deltas)\cos(\deltap)},~
    Y^{(i)}_{r\cos(\deltas)\cos(\deltap)},~
    Y^{(i)}_{\sin(\deltas)\cos(\deltas)\cos(\deltap)},~
    Y^{(i)}_{\cos(\deltas)\sin(\deltap)\cos(\deltap)},~
    Y^{(i)}_{uv\subm \np},~
    Y^{(i)}_{ur\np},~
    Y^{(i)}_{u\sin(\deltas)\np},~
    Y^{(i)}_{u\sin(\deltap)\np},~
    Y^{(i)}_{v\subm \cos(\deltas)\np},~
    Y^{(i)}_{v\subm \cos(\deltap)\np},~
    Y^{(i)}_{v\subm \np\np},~
    Y^{(i)}_{r\cos(\deltas)\np},~
        Y^{(i)}_{r\cos(\deltap)\np},~
        Y^{(i)}_{r\np\np},~
        Y^{(i)}_{\sin(\deltas)\cos(\deltas)\np},~
        Y^{(i)}_{\sin(\deltas)\cos(\deltap)\np},~
    Y^{(i)}_{\sin(\deltas)\np\np},~
    Y^{(i)}_{\cos(\deltas)\sin(\deltap)\np},~
    Y^{(i)}_{\sin(\deltap)\cos(\deltap)\np},~
    Y^{(i)}_{\sin(\deltap)\np\np} \big)

%% file: z_odd.tex
\boldsymbol{z}_{\text{odd}} =  
\big(
v\subm ,~r,~\sin(\deltas),~\sin(\deltap),~
uv\subm ,~
ur,~
u\sin(\deltas),~
u\sin(\deltap),~
v\subm \cos(\deltas),~
v\subm \cos(\deltap),~
v\subm \np,~
r\cos(\deltas),~
r\cos(\deltap),~
r\np,~
\sin(\deltas)\cos(\deltas),~
\sin(\deltas)\cos(\deltap),~
\sin(\deltas)\np,~
\cos(\deltas)\sin(\deltap),~
\sin(\deltap)\cos(\deltap),~
\sin(\deltap)\np,~
uuv\subm ,~v\subm v\subm v\subm ,~uur,~v\subm |v\subm |r,~v\subm r|r|,~
rrr,~
uu\sin(\deltas),~
v\subm |v\subm |\sin(\deltas),~
v\subm r\sin(\deltas),~
r|r|\sin(\deltas),~
uv\subm \cos(\deltas),~
ur\cos(\deltas),~
u\sin(\deltas)\cos(\deltas),~
v\subm \cos(\deltas)\cos(\deltas),~
r\cos(\deltas)\cos(\deltas),~
\sin(\deltas)\cos(\deltas)\cos(\deltas),~
uu\sin(\deltap),~
u\cos(\deltas)\sin(\deltap),~
v\subm |v\subm |\sin(\deltap),~
v\subm r\sin(\deltap),~
v\subm \sin(\deltas)\sin(\deltap),~
r|r|\sin(\deltap),~
r\sin(\deltas)\sin(\deltap),~
\cos(\deltas)\cos(\deltas)\sin(\deltap),~
uv\subm \cos(\deltap),~
ur\cos(\deltap),~
u\sin(\deltas)\cos(\deltap),~
u\sin(\deltap)\cos(\deltap),~
v\subm \cos(\deltas)\cos(\deltap),~
r\cos(\deltas)\cos(\deltap),~
\sin(\deltas)\cos(\deltas)\cos(\deltap),~
\cos(\deltas)\sin(\deltap)\cos(\deltap),~
uv\subm \np,~
ur\np,~
u\sin(\deltas)\np,~
u\sin(\deltap)\np,~
v\subm \cos(\deltas)\np,~
v\subm \cos(\deltap)\np,~
v\subm \np\np,~
r\cos(\deltas)\np,~
r\cos(\deltap)\np,~
r\np\np,~
\sin(\deltas)\cos(\deltas)\np,~
\sin(\deltas)\cos(\deltap)\np,~
\sin(\deltas)\np\np,~
\cos(\deltas)\sin(\deltap)\np,~
\sin(\deltap)\cos(\deltap)\np,~
\sin(\deltap)\np\np \big) \enspace .